\documentclass[aps,prb,longbibliography,floatfix,reprint]{revtex4-1}
\usepackage[normalem]{ulem}
\usepackage{cancel}
\usepackage{amsmath}
\usepackage{graphicx}
\usepackage{amsfonts}
\usepackage{color}
\usepackage{placeins}
\newcommand{\stkout}[1]{\ifmmode\text{\sout{\ensuremath{#1}}}\else\sout{#1}\fi}

\begin{document}

\title{Quasiparticles and phonon satellites in spectral functions of semiconductors and insulators:
Cumulants applied to full first principles theory and Fr\"ohlich polaron.}

\author{ Jean Paul Nery }
\email{jeanpaul240@gmail.com}
\affiliation{ Department of Physics and Astronomy,
              Stony Brook University,
              Stony Brook, New York 11794-3800, USA }

\author{ Philip B. Allen }
\email{philip.allen@stonybrook.edu}
\affiliation{ Department of Physics and Astronomy,
              Stony Brook University, 
              Stony Brook, New York 11794-3800, USA }

\author{ Gabriel Antonius }
\email{antonius@lbl.gov}
\affiliation{Department of Physics, University of California at Berkeley, California 94720, USA 
and Materials Sciences Division, Lawrence Berkeley National Laboratory, Berkeley, California 94720, USA} 

\author{ Lucia Reining}
\email{lucia.reining@polytechnique.fr}
\affiliation{ Laboratoire des Solides Irradi\'es,
              Ecole Polytechnique, CNRS, CEA/DSM and European Theoretical Spectroscopy Facility (ETSF),
              91128 Palaiseau, France}

\author{ Anna Miglio }
\email{anna.miglio@uclouvain.be}
\affiliation{Institute of Condensed Matter and Nanoscience, European Theoretical Spectroscopy Facility (ETSF), Universit\'e Catholique de Louvain,
B-1348 Louvain-la-Neuve, Belgium} 

\author{ Xavier Gonze }
\email{xavier.gonze@uclouvain.be}
\affiliation{Institute of Condensed Matter and Nanoscience, European Theoretical Spectroscopy Facility (ETSF), Universit\'e Catholique de Louvain,
B-1348 Louvain-la-Neuve, Belgium} 

\date{\today}

\begin{abstract}

The electron-phonon interaction causes thermal and zero-point motion shifts of 
electron quasiparticle (QP) energies $\epsilon_k(T)$.  Other consequences of interactions,
visible in angle-resolved photoemission spectroscopy (ARPES) experiments,
are broadening of QP peaks and appearance of sidebands, contained in
the electron spectral function 
 $A(k,\omega)=-{\Im m}G_R(k,\omega) /\pi$, where
$G_R$ is the retarded Green's function. 
Electronic structure codes ({\it e.g.} using density-functional
theory) are now available that compute the shifts and start to address
broadening and sidebands. Here we consider MgO and LiF, and determine their
nonadiabatic Migdal self energy. The spectral function obtained from
the Dyson equation makes errors in the weight and energy of the QP 
peak and the position and weight of the phonon-induced sidebands.
Only one phonon satellite appears, 
with an unphysically large energy difference (larger than the highest phonon energy) with respect to the QP peak.  
By contrast, the spectral function from a cumulant treatment of the same self energy
is physically better, giving a quite accurate QP energy and several satellites 
approximately spaced by the LO phonon energy.
In particular, the positions of the QP peak and first satellite agree closely with those found for the 
Fr\"ohlich Hamiltonian by Mishchenko {\it et al.} (2000) using diagrammatic Monte Carlo. 
We provide a detailed comparison between the first-principles MgO and LiF 
results and those of the Fr\"ohlich Hamiltonian. Such an analysis applies widely
to materials with infra-red active phonons. 
We also compare the retarded and time-ordered cumulant treatments: 
they are equivalent for the Fr\"ohlich Hamiltonian, and only 
slightly differ in first-principles electron-phonon results for wide-band gap materials.

\end{abstract}

\maketitle

%***************************************************************************
\section{Introduction} \label{sec:intro}

The notion of a single-particle (quasiparticle, or QP) spectrum $\epsilon_k$ for electrons
($k$ is short for all needed quantum numbers - wavevector, band, spin - 
$\mathbf{k},n,\sigma$) underlies much of 
solid state physics.  Justification of the existence of such quasiparticle spectrum 
\cite{Landau1956,Martin2016} relies on experiment.  Optical experiments, combined with 
theoretical guidance, {\it e.g.} the ``empirical pseudopotential method'' \cite{Cohen1966}, 
have been used for decades, and have allowed extraction of
accurate $\epsilon_k$ from reflectivity data for simple semiconductors.  Excitonic effects
cause deviation from an independent-particle interpretation, but theory 
can determine their consequences and help to extract one-electron properties. 
Angle-resolved
photoemission spectroscopy (ARPES) provides a more direct QP
spectrum \cite{Berglund1964,Koyama1970}.   The data can be approximately related to the
rigorously defined one-particle spectral function, obtained from the
retarded Green's function $G(k,\omega)$ as
\cite{Altland2010,Martin2016}
\begin{equation}
A(k,\omega)=-\frac{1}{\pi}{\Im m} G_R(k,\omega),
\label{eq:A}
\end{equation}
where $G_R(k,\omega)$ is the Fourier transform 
\begin{equation}
\int dt \exp(i\omega t)G_R(k,t)
\end{equation}
of the retarded Green's function
\begin{equation}
G_R(k,t)=-i\langle\{c_k(t),c_k^+(0)\}\rangle\theta(t),
\label{eq:GR}
\end{equation}
and $\{a,b\}$ is an anticommutator.
At temperature $T=0$, the time-ordered (t-O) Green's function 
yields also a simple expression,
\begin{equation}
A(k,\omega)=|{\Im m} G_{t-O}(k,\omega)|/\pi,
\end{equation}
since the $T=0$ t-O function has the same 
imaginary part as $G_R(k,\omega)$, except for a sign change at $\omega=\mu$. 

When the spectral function $A(k,\omega)$ exhibits a strong peak
that correlates with the corresponding single-particle theory, this defines a QP energy.
The total spectral weight $\int d\omega A(k,\omega)$
is 1, but the QP peak has reduced weight $Z_{k}<1$.  It is
broadened, and accompanied by features at other energies.  
When clearly differing from a structureless background, these features are called 
satellites.  This has been seen in many photoemission experiments,
for example, Ref.~\onlinecite{Moser2013}
for polaron satellites, and Ref.~\onlinecite{Steiner1979} for plasmon satellites.
Thus $A(k,\omega)$ contains more information than just
the QP energy $\epsilon_k$. 
Full interpretation is a challenge to theory; progress on plasmon \cite{Aryasetiawan1996,Guzzo2011}
and polaron \cite{Verdi2017} satellites in real semiconductors is occuring. 

On the basis of a one-band analysis for metals, Migdal \cite{Migdal1958} showed that the electron self energy due to phonons
has important low temperature dynamical effects, which can be accurately described by a lowest-order
self energy diagram $\Sigma_M$.  Vertex corrections can be omitted because of the small ratio $\omega_{\rm ph}/E_F$
of phonon energies to the Fermi energy.  Even though this argument that allows to neglect vertex corrections does not strictly apply in semiconductors, we
will use the term ``Migdal'' to indicate the Migdal formula, $\Sigma_M$, given later as Eq.(\ref{self_energy_Fan}),
where it is labeled $\Sigma^{\rm Fan}$, refering to earlier works on insulators by Fan 
in the fifties\cite{Fan1950,Fan1951}.  Using Dyson's equation, the corresponding Green's function is
\begin{equation}
G_D(k,\omega)=G_0(k,\omega)+G_0(k,\omega)\Sigma_M(k,\omega)G_D(k,\omega).
\label{eq:GD}
\end{equation}
When inserted into Eq.\eqref{eq:A}, it gives a spectral function which we will call ``Dyson-Migdal'', $A_{\rm DM}$.  
Given the success of Migdal theory in metals and the expected success of low order perturbation theory
for most electron-phonon problems, it is embarassing to realize \cite{Dunn1975,Gunnarsson1994,Story2014} 
that $A_{\rm DM}(k,\omega)$ often agrees poorly 
with measured $A(k,\omega)$.  Typically, only one distinct satellite is found 
(somewhat misplaced) in the side-band spectrum, while in reality
several satellites are possible \cite{Moser2013}, corresponding to  emission of several phonons.
Even worse, the QP peak is misplaced \cite{note-selfconsistency}.

An alternative approach, also approximate, involves a cumulant treatment (see e.g. 
Refs.~\onlinecite{Mahan2000,Almbladh1983,Hedin1999, Gumhalter2016} and pp. 410--415 of Ref.~\onlinecite{Martin2016}).
This will be denoted as $G_C$ for the Green's function, and
$A_{\rm C}$ for the spectral function.  
The cumulant was advocated for describing plasmon satellite effects
in metals by Steiner {\it et al.} \cite{Steiner1979}
and Hedin \cite{Hedin1980}.  For electron-phonon effects, the earliest use was by Dunn \cite{Dunn1975} 
for the polaron spectral function.  Gunnarsson {\it et al.} \cite{Gunnarsson1994} introduced 
the cumulant method for electron-phonon effects in metals.  Verdi {\it et al.} \cite{Verdi2017}
applied the cumulant method to doped semiconductors, where both plasmon and phonon effects
occur.  Here we apply the cumulant method to undoped semiconductors 
with electron-phonon renormalization but no dynamical electron-electron coupling.

\begin{figure}
\includegraphics[width=0.45\textwidth]{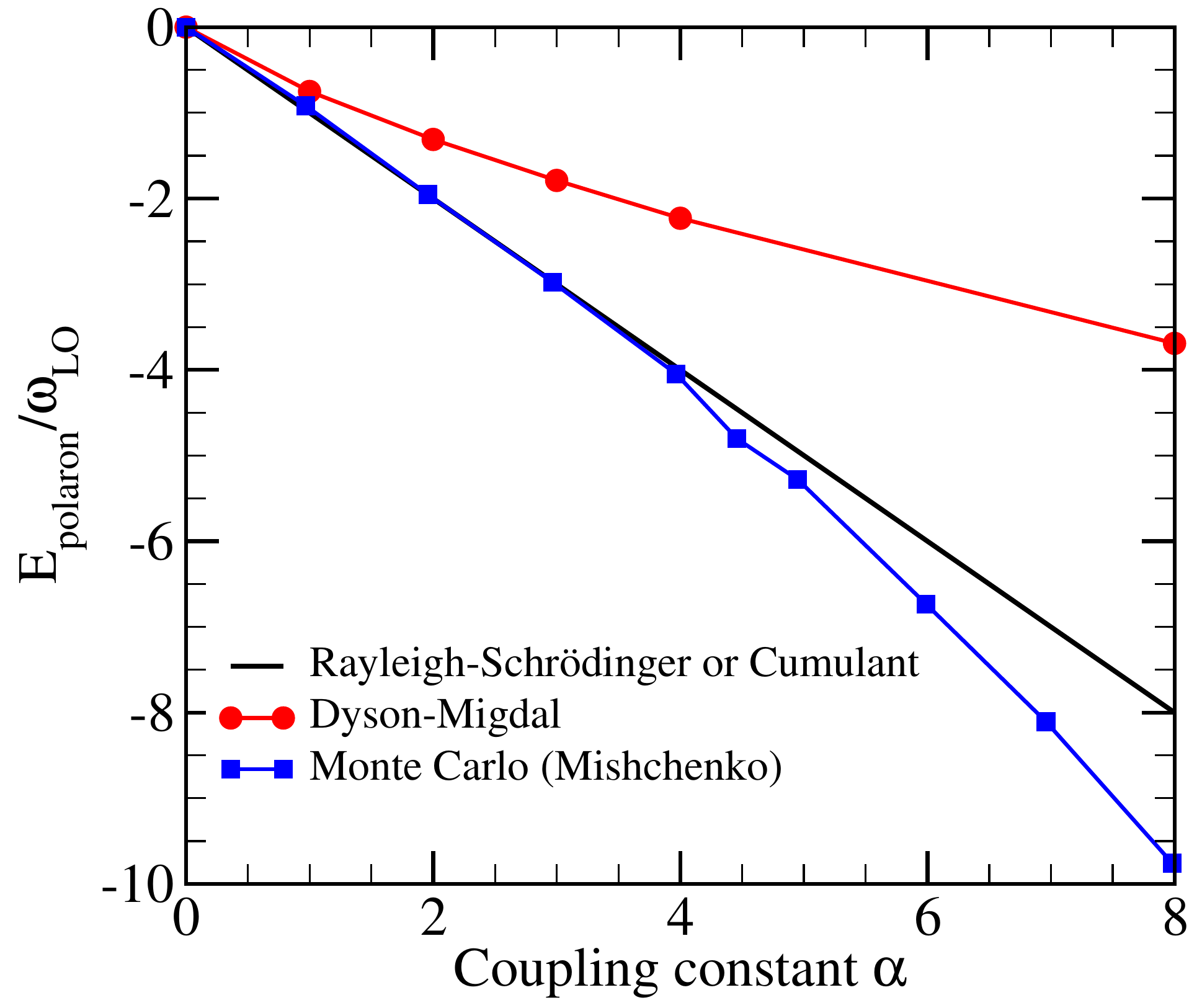}
\caption{Quasiparticle energy of the Fr\"ohlich polaron, as a function of the Fr\"ohlich coupling constant, 
in units of $\omega_{LO}$: 
accurate results from Ref.~\onlinecite{Mishchenko2000} (blue squares), from the cumulant
approach (which agrees with lowest-order 
Rayleigh-Schr\"odinger theory) (black line),
and from Dyson-Migdal spectral function \cite{note-selfconsistency,Mahan2000} (red circles).
 \label{fig:QPenergy}}
\end{figure}

To further motivate the interest to work with the cumulant approach instead of the Dyson-Migdal one, 
consider Fig. \ref{fig:QPenergy}, which illustrates where the QP peak appears in 
various treatments of the Fr\"ohlich polaron problem \cite{Froehlich1954,Mahan2000}.  
Lowest-order Rayleigh-Schr\"odinger perturbation
theory puts the renormalized polaron energy below the conduction band minimum by $\alpha\omega_{LO}$,
where $\alpha$ is the Fr\"ohlich coupling constant.  An accurate high-order treatment
\cite{Mishchenko2000} gives a surprisingly similar answer, shown as the blue squares on the graph.  This is
where the QP peak of the spectral function should be located.  
If the Migdal self energy $\Sigma_M$ is evaluated at $\omega=\epsilon_k$, the unperturbed energy,
the corresponding energy shift $\Delta \epsilon_{k=0}=\Sigma_M(0,\epsilon_k)$ agrees with the
Rayleigh-Schr\"odinger result $-\alpha\omega_{LO}$.     
However, the actual QP peak of the spectral function $A_{\rm DM}$ is located at a fairly different energy, shown by the red circles on the graph\cite{note-selfconsistency,Mahan2000}.  The cumulant method used in the present paper puts the QP
peak exactly on the Rayleigh-Schr\"odinger line, close to the correct polaron answer. 
The cumulant method also greatly
improves the position of the first satellite peak, as will be shown later.
It also gives additional peaks, corresponding to multiphonon excitations. However, they
differ in location and strength from the other peaks in the accurate Fr\"ohlich spectral function
\cite{Mishchenko2000}.

\begin{figure}
\includegraphics[width=0.45\textwidth]{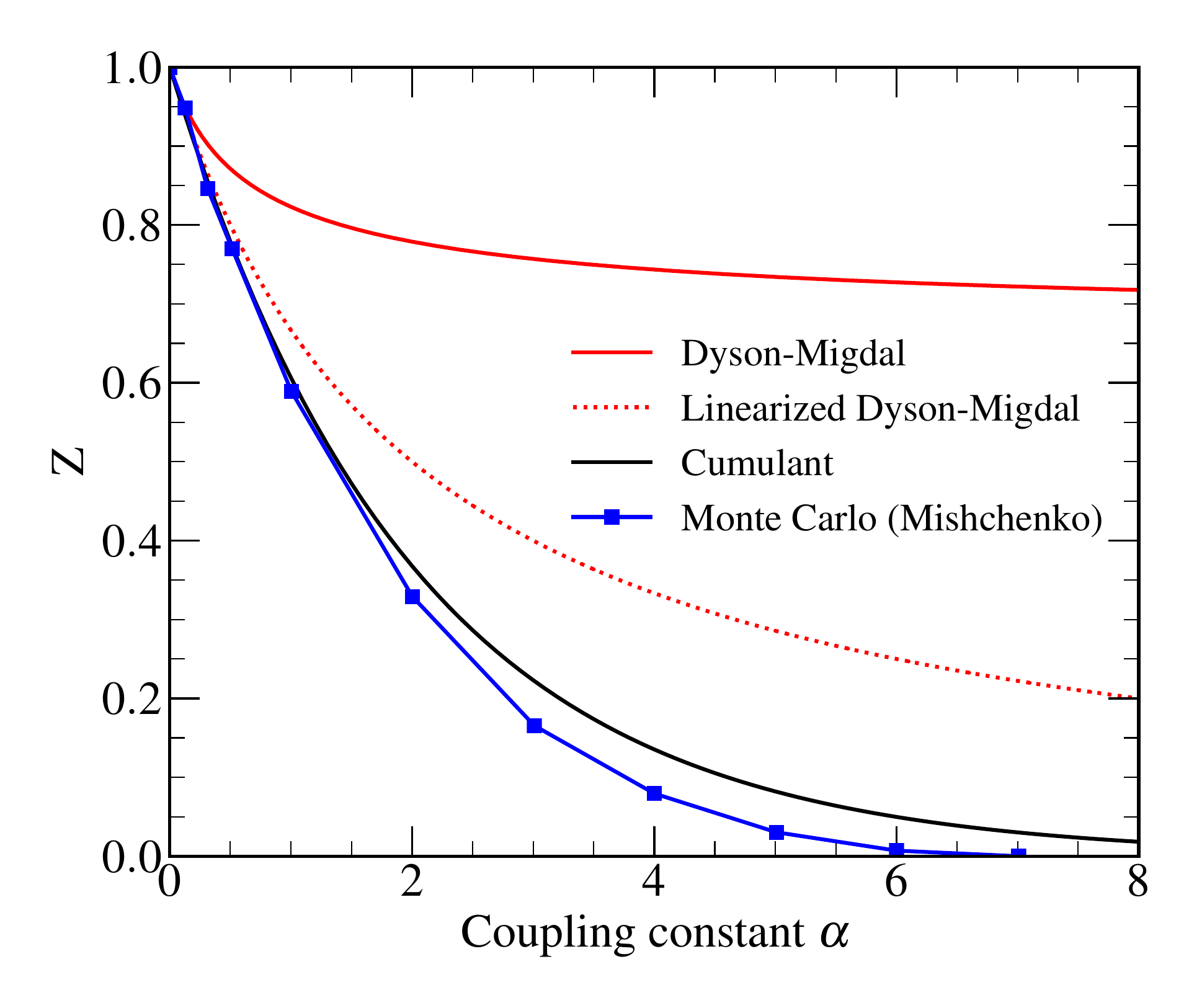}
\caption{Quasiparticle spectral weight $Z_{k=0}$ of the Fr\"ohlich polaron, as a function of the 
Fr\"ohlich coupling constant. 
The blue squares are accurate Monte Carlo results from Ref.~\onlinecite{Mishchenko2000}. 
The solid red line comes from the full Dyson-Migdal spectral function, 
Eq. \eqref{eq:ZDM}.
The dotted red line comes from the linearized approach, Eq.(\ref{eq:ZDMlin}).
The black line is the cumulant result using Eq. \eqref{eq:ZR} with the retarded Green's function. 
 \label{fig:QPZ}}
\end{figure}

Another improvement given by the cumulant method is a much more reasonable value of the QP
spectral weight $Z$, defined later in Eq.(\ref{eq:ZDM}) and shown in Fig. \ref{fig:QPZ}.


Since the characteristic energy of plasmons is much bigger than that
of phonons, the resolution needed to see them in photoemission experiments
has been available for many years.  The valence photoemission spectra of alkali metals,
that exhibit multiple plasmon satellite structures, was modeled from first principles by
Aryasetiawan {\it et al.} \cite{Aryasetiawan1996} using a cumulant treatment.
More recently, Kheifets {\it et al.} \cite{Kheifets2003} and Guzzo {\it et al.} \cite{Guzzo2011} 
did ARPES for
valence electrons in Si.  They observed plasmon satellites caused by screened Coulomb
interaction. Their calculated spectrum based on a cumulant approach $G_C$, agrees much better with their data
than a theory based on the Dyson $G_D$.  
Similar observations were made for graphene, 
doped graphene and graphite\cite{Lischner2013, Guzzo2014}, and for several other materials
\cite{Mahan2000b, Steiner1979}.
The homogeneous electron gas has also been studied using the cumulant approach
\cite{Holm1997,Caruso2016,Vigil-Fowler2016}.

Electron-phonon interaction (EPI) effects are ubiquitous in solids \cite{Giustino2017}, 
but were not seen in ARPES until improved
resolution became available.  In metals, temperature shifts of photoemission linewidths
were seen for surface states in Cu \cite{McDougall1995} and Ag \cite{Eiguren2002}. 
Phonon-induced ``kinks'' in the quasiparticle
dispersion were seen by photoemission from surface states \cite{Plummer2003} 
of Be(0001) \cite{Hengsberger1999,LaShell2000} and Mo(110) \cite{Valla1999}.
Bulk electron-phonon effects were seen by ARPES in superconducting 
and normal Pb \cite{Reinert2003}.  None of these experiments on metals have resolved 
EPI-induced satellite structures.
ARPES studies of EPI effects on electrons near the band gaps of semiconductors are now available,
with resolved EPI-induced satellites for electrons
and holes doped into non-metals, for bulk \cite{Moser2013},
surface \cite{Chen2015,Yukawa2016}, and interface \cite{Cancellieri2016} doped regions.  

First-principle studies of the EPI effects on the electronic structure of insulators (zero-point motion as well as 
temperature dependence) have mostly used perturbation theory to second order
\cite{Allen1976}, including both second-order
corrections from the first-order matrix element (Fan terms, the same correction
that Migdal deals with, but often treated adiabatically, and including virtual interband transitions) 
and first-order effects from the second-order
matrix elements (Debye-Waller terms).
 The focus has mostly been on the quasiparticle shift in simple solids,
and occasionally on the quasiparticle broadening \cite{Marini2008,Giustino2010,Cannuccia2011,Cannuccia2012, 
Monserrat2013, Kawai2014, Monserrat2014a, Monserrat2014b,
Ponce2014a,Antonius2014,Ponce2014b,Monserrat2015,Ponce2015,Antonius2015, Friedrich2015, Monserrat2016, 
Molina2016,Nery2016,Villegas2016,Saidi2016,Giustino2017}. 
Polarons, by contrast, have been studied for a long time using high-order perturbative treatment of the
singular Fr\"ohlich first-order matrix element, with no second or higher-order matrix elements, and 
simplified model band structures \cite{Mahan2000,Devreese2009}.  
When the Fr\"ohlich coupling constant $\alpha$ exceeds $\approx$ 5, nonperturbative corrections
start to be needed, and eventually small polarons may form \cite{Emin2012} in real materials
where the polaron radius in the continuum approach would be comparable or smaller than the interatomic distance.  
Also, defects or higher than harmonic lattice displacement terms cannot be ignored in real materials.
Only recently \cite{Ponce2015} 
was it noticed that first-principles studies of semiconductors had incorrectly ignored nonadiabatic aspects
familiar in polaron literature for the quasiparticle energy shift (also sometimes referred to as  renormalization) due to LO phonons in infrared-active semiconductors.  
Perturbation theory had previously been simplified by omission of $\pm \omega_Q$ from
denominators.  Because of the $1/q$ divergence of the Fr\"ohlich coupling,
LO phonon contributions cause unphysical divergences (3D-integral of $1/q^4$ at small $q$ phonon wavevectors)  
if the $\pm\omega_Q$ pieces are omitted \cite{Ponce2015}.

The literature on large polarons has focussed on models such as the Fr\"ohlich Hamiltonian, 
generally ignoring the existence of multiple phonon branches, non-parabolic electronic bands, Debye-Waller,  and interband effects.   
We advocate unification of the separate skills of polaron and energy-band communities. 
This has started, with the above-mentioned
understanding of the LO-phonon role in first-principles calculations \cite{Ponce2015,Nery2016,Lambrecht2017}, 
as well as the first-principles approach to the Fr\"ohlich vertex developed by Sjakste {\it et al.} \cite{Sjakste2015}
and Verdi and Giustino \cite{Verdi2015}.

Beyond the computation of quasiparticle shifts, first-principles studies of spectral functions, side bands
and satellites have also appeared. 
Cannuccia and Marini \cite{Cannuccia2011, Cannuccia2012} showed that optical data for diamond and polyethylene
contain subgap EPI effects also visible in the single-electron theoretical spectral function $A(k,\omega)$.
Spectral functions were computed by electron-phonon perturbation theory for the full band structure of
C, BN, MgO and LiF by Antonius {\it et al.} \cite{Antonius2015}.
A satellite is distinctly seen in the spectral function at the top of the valence band of LiF.
However, these computations used $G_D$ with a Migdal self-energy.  
Cumulant studies of EPI effects \cite{Dunn1975} have recently been revived by Verdi {\it et al.} \cite{Verdi2017}
who discussed the doped TiO$_2$ data of Ref.~\onlinecite{Moser2013}, explaining
the evolution from polaron to metallic-type EPI-renormalization as the doping level increases.
The current paper reconsiders the results of Antonius {\it et al.} \cite{Antonius2015} using $G_C$,
and shows, on the basis of the comparison with the Fr\"ohlich Hamiltonian, how a cumulant calculation
improves such results.

Cumulants \cite{Kubo1962}, and their relatives, have been discussed a lot in older 
\cite{Langreth1970,Dunn1975,Hedin1980,Hedin1991,Gunnarsson1994,Aryasetiawan1996,Hedin1999}
as well as more recent \cite{Guzzo2011,Lischner2013,Kas2014,Story2014,Zhou2015,
Gumhalter2016,Mayers2016,Martin2016,Vigil-Fowler2016} literature.  The improvement they give is
incomplete \cite{Gumhalter2016,McClain2016,Mayers2016}. 
One issue is whether the Green's function $G$ should be time-ordered (t-O)
(as in Refs.~\onlinecite{Aryasetiawan1996,Guzzo2011}) or retarded (R) (advocated by Kas, Rehr, and Reining \cite{Kas2014}, and used for the electron-phonon interaction in the case of metallic systems
in Ref.~\onlinecite{Story2014}). An exact many-body theory can be formulated
based on either retarded Green's functions $G_R$, or t-O Green's functions, $G_{t-O}$.
However, approximate calculations, {\it e.g.} a cumulant treatment based on low-order 
perturbation theory, may yield different results when different
versions, $G_R$ or $G_{t-O}$, are used.
In diagrammatic perturbation theory, the t-O Green's function is usually preferred
at $T=0$, while the retarded Green's function enters more naturally at $T>0$, 
through analytic continuation of Matsubara Green's functions to the real frequency axis \cite{Martin2016}.  
Both versions of $G$ give very similar cumulant spectral functions $A$ in the case of a single electronic band, and 
are expected to yield minor differences if the occupied and unoccupied bands differ by a gap large compared to the 
energy of the excitations appearing in the self energy. 
By contrast, for the homogeneous electron gas, they show clear differences \cite{Kas2014}. The different cumulant expansions for real systems have
been examined in depth in Ref.~\onlinecite{ZhouThesis2015}.

Quantitative tests for EPI effects in non-metals treated with a cumulant approach are almost nonexistent. 
We are only aware of the above-mentioned
model calculations for the Fr\"ohlich Hamiltonian by Dunn \cite{Dunn1975},
and first-principles study of TiO$_2$ by  Verdi {\it et al.} \cite{Verdi2017}.

The current paper deals with the zero-temperature spectral function $A(k,\omega,T=0)$, 
for states $k$ at band extrema near the gap of two real insulators, MgO and LiF, 
arising from EPI effects.
The Migdal method is 
used for $\Sigma(k,\omega)$
including, for polar EPI's, the $\pm\omega_Q$ in energy denominators, see Eqs. (\ref{self_energy_DW+Fan}-\ref{self_energy_Fan_oc}).  
The Migdal level of perturbation theory assumes 
that the Fr\"ohlich coupling constant $\alpha$ is not too big.  We compute the spectral function both at the
Dyson-Migdal ($A_{\rm DM}(k,\omega)$) and cumulant ($A_{\rm C}(k,\omega)$) levels, and show that the
cumulant version is physically more sensible. 

We first summarize the theoretical background for (a) computation of self energies, and (b) spectral functions.
Then we consider the Fr\"ohlich Hamiltonian, and discuss results for several values of the coupling constant $\alpha$. 
Subsequently we calculate the self energy and the spectral function,
for the top of the valence band and bottom of conduction band of MgO and LiF, 
using {\it ab initio} density functional perturbation theory (DFPT) calculations with the code ABINIT \cite{Gonze2009,Gonze2016} to determine phonons and their coupling with electrons. 
We consider, for these states, the Dyson-Migdal (D-M) approach, as well as 
different flavors of the cumulant approach. Given
the small phonon frequencies with respect to the wide gap of these materials, the results are essentially identical for the
different cumulant approaches, while the D-M approach gives qualitatively very different results.
This calls for a reconsideration of the results based on the D-M approach 
given in Refs.~\onlinecite{Cannuccia2011,Cannuccia2012,Antonius2015}.

\section{SELF-ENERGY}

The Hartree atomic unit system is used throughout ($\hbar=m_e=e=1$).
Starting from now, we will use the more explicit notations $\mathbf{k}$ for wavevectors and $n$ for bands, instead of $k$. Spin is irrelevant in this article.
The self energy of an unperturbed state, labelled by wavevector and band, 
includes two contributions at the lowest order of perturbation theory (quadratic in the strength of the EPI),
namely, the Fan self energy and the Debye-Waller self energy \cite{Giustino2017}:
\begin{eqnarray}
\Sigma(\mathbf{k}n,\omega) = \Sigma^{\mathrm{Fan}}(\mathbf{k}n,\omega) + \Sigma^{\mathrm{DW}}(\mathbf{k}n).
\label{self_energy_DW+Fan}
\end{eqnarray}

The Debye-Waller self energy (see its expression in Refs.~\onlinecite{Ponce2014a,Giustino2017}) is static and real. 
On the contrary, the Fan self energy is dynamical.
In matrix notation, the Fan self energy is given by $\Sigma = i G \Gamma D$, 
where $G$ is the electron propagator, 
$\Gamma$ is the vertex, and $D$ is the phonon propagator. 
As often done \cite{Migdal1958}, the vertex is approximated as $\Gamma=1$. 
Approximating the electronic Green's function by its non-interacting counterpart, $G=G^{(0)}$, e.g. Kohn-Sham Green's function without electron-phonon corrections,  corresponds to a non-self-consistent treatment, 
and gives the standard result for the retarded Fan self energy \cite{Giustino2017} :
\begin{eqnarray}
&&\Sigma^{\mathrm{Fan}}(\mathbf{k}n,\omega) = \frac{1}{N_{\mathbf{q}}} \sum_{\mathbf{q}j}^{\rm BZ} \sum_{n'} |\langle \mathbf{k}n|H^{(1)}_{\mathbf{q}j}|\mathbf{k+q}n'\rangle|^2 \times \nonumber \\
&&\left[ 
\frac{n_{\mathbf{q}j} + 1-f_{\mathbf{k+q}n'}}{\omega-\varepsilon_{\mathbf{k+q}n'}-
\omega_{\mathbf{q}j}+i \eta}+ 
\frac{n_{\mathbf{q}j}+f_{\mathbf{k+q}n'}}{\omega-\varepsilon_{\mathbf{k+q}n'}+
  \omega_{\textbf{q}j}+i \eta} \right].
\nonumber
\\
\label{self_energy_Fan}
\end{eqnarray}
In this expression, contributions from phonon modes with harmonic
%non-interacting -- no, I regard DFPT phonons as "interacting" but harmonic (PBA)
phonon energy $\omega_{\mathbf{q}j}$ and occupation number $n_{\mathbf{q}j}$,
are summed for all phonon branches, labelled $j$, and wavevectors, labelled $\mathbf{q}$, in the entire Brillouin zone.
The latter is discretized, with $N_{\mathbf{q}}$
indicating the number of wavevectors in the sum. The limit for infinite number of wavevectors is implied.
Contributions from intermediate electronic states $|\mathbf{k+q}n'\rangle$ with 
%non-interacting -- ?
electron energy $\varepsilon_{\mathbf{k+q}n'}$ (not renormalized by phonons)
and occupation number $f_{\mathbf{k+q}n'}$
are summed for all bands $n'$ (valence and conduction). 
$H^{(1)}_{\mathbf{q}j}$ is the self-consistent change of potential due to the phonon labelled $\mathbf{q}j$ \cite{Giustino2017,Marini2015}.  Eq.(\ref{self_energy_Fan}) is also the Migdal result $\Sigma_M$.

The t-O Fan self energy is obtained from the above retarded self energy by multiplying the
infinitesimal quantity $\eta$ by $\rm{sign}(\varepsilon_{\mathbf{k+q}n'}+\omega_{\mathbf{q}j}-\mu)$ in the first denominator, 
and by  $\rm{sign}(\varepsilon_{\mathbf{k+q}n'}-\omega_{\mathbf{q}j}-\mu)$  in the second denominator.
The retarded self energy has all poles below the real axis, while the poles of the t-O self energy
are below the real axis if the pole is above the chemical potential energy $\mu$, and above the real axis if the pole
is below the chemical potential energy $\mu$.

In the present article, we will work with semiconductors at zero temperature, 
in which case the phonon occupation numbers vanish, 
and the electron occupation numbers $f_{\mathbf{k+q}n'}$ are either one, for the valence states, or zero, for the conduction states. 
There are separate contributions to the Fan self energy
from the intermediate states $|\mathbf{k+q}n'\rangle$ in the conduction bands (labelled `un' for unoccupied) 
and in the valence band (labelled `oc' for occupied) \cite{notation-splitting}. No matter whether the initial state $|\mathbf{k}n\rangle$
is from the valence or the conduction band, both contributions occur,
\begin{equation}
\Sigma^{\mathrm{Fan}}(\mathbf{k}n,\omega)= \Sigma_{un}^{\mathrm{Fan}}(\mathbf{k}n,\omega)
+\Sigma_{oc}^{\mathrm{Fan}}(\mathbf{k}n,\omega). 
\label{eq:self_energy_Fan_e+h}
\end{equation}
Explicit equations are
\begin{eqnarray}
\Sigma_{un}^{\mathrm{Fan}}(\mathbf{k}n,\omega) = \frac{1}{N_{\mathbf{q}}} \sum_{\mathbf{q}j}^{\rm BZ} 
\sum_{n'}^{unocc} \frac{ |\langle \mathbf{k}n|H^{(1)}_j|\mathbf{k+q}n'\rangle|^2 }
{ \omega-\varepsilon_{\mathbf{k+q}n'}-\omega_{\mathbf{q}j}+i \eta },\nonumber
\\
\label{self_energy_Fan_un}
\end{eqnarray}
for the intermediate unoccupied state contribution to the self energy (be it retarded or t-O), and
\begin{eqnarray}
\Sigma_{oc}^{\mathrm{Fan}}(\mathbf{k}n,\omega) = \frac{1}{N_{\mathbf{q}}} 
\sum_{\mathbf{q}j}^{\rm BZ} \sum_{n'}^{occ}
\frac{ |\langle \mathbf{k}n|H^{(1)}_j|\mathbf{k+q}n'\rangle|^2 }
{ \omega-\varepsilon_{\mathbf{k+q}n'}+\omega_{\mathbf{q}j}+i \eta },\nonumber
\\
\label{self_energy_Fan_oc}
\end{eqnarray}
for the intermediate occupied state contribution to the retarded self energy (for the t-O one, the sign of $i\eta$ is changed,
which changes the imaginary part of the self energy, but not the real part).
The imaginary part of the retarded self energy is always negative, while the t-O one is negative above the chemical potential,
and positive below it. Only the retarded self energy satisfies Kramers-Kronig relationships.

For semiconductors with infrared-active phonons, the intraband ($n^\prime =n$) contribution with 
small-$q$ LO phonons (the Fr\"ohlich problem) gives the most important dynamical features 
for the frequency range near the 
bare electronic energy ($\omega\approx\epsilon_{\mathbf{k}n}$), due to the combination of  small-$q$ 
diverging matrix element
$\langle \mathbf{k}n|H^{(1)}_{\mathbf{q}j}|\mathbf{k+q}n'\rangle \rightarrow C_n \delta_{nn'} /q$, see Ref.~\onlinecite{Ponce2015},
and small denominator ($\omega\approx\epsilon_{\mathbf{k}n}\approx\epsilon_{\mathbf{k+q}n}$)  
in Eqs.(\ref{self_energy_Fan_un})-(\ref{self_energy_Fan_oc}). 

\section{QUASIPARTICLES AND SPECTRAL FUNCTION}\label{sec:QP}

The second-order self energy $\Sigma(\mathbf{k}n,\omega)$ is the basis 
of different approximations for the quasiparticle shift 
and the spectral function.  In the Rayleigh-Schr\"odinger (RS)
approximation, the new quasiparticle energy $E_{\mathbf{k}n}$ is shifted from its initial value $\epsilon_{\mathbf{k}n}$
by the real part of the self energy evaluated at $\epsilon_{\mathbf{k}n}$:   
\begin{equation}
E^{RS}_{\mathbf{k}n}= \epsilon_{\mathbf{k}n}
+\Re e \Sigma(\mathbf{k}n,\omega=\epsilon_{\mathbf{k}n}).
\label{eq:ERS}
\end{equation}
The spectral function $|{\Im m}G(\mathbf{k}n,\omega)|/\pi$
(either t-O or retarded $G$, since we take $T=0$) has
dynamical effects beyond Rayleigh-Schr\"odinger. 
%(i) There is a shift of the quasiparticle peak away from $E^{RS}$, explained below.
%(ii) A phonon-emission side band appears at a shifted energy.
%(iii) Except at the bottom of the conduction band or at the top of the valence band
%(and at $T=0$) the quasiparticle peak broadens.
%
In the Dyson-Migdal (D-M) approach \cite{Giustino2017}, the spectral function obtained from the self energy is
\begin{eqnarray}
A_{\rm DM}(\mathbf{k}n,\omega) &=& \frac{1}{\pi} | {\Im m} G_D(\mathbf{k}n,\omega) | \nonumber  \\
&=& %\frac{1}{\pi}
\frac{(1/\pi) |\Im m \Sigma(\mathbf{k}n,\omega)| }
       {(\omega-\varepsilon_{\mathbf{k}n}-\Re e \Sigma(\mathbf{k}n,\omega))^2+({\Im m \Sigma(\mathbf{k}n,\omega))^2 }}. 
   \nonumber \\ 
\label{eq:Spectral_DM}
\end{eqnarray}
There is typically a QP peak at $\omega=E^D_{\mathbf{k}n}$ where
$\omega- \epsilon_{\mathbf{k}n}+\Re e \Sigma(\mathbf{k}n,\omega)=0$
({\it i.e.} where the first term in the denominator of Eq.(\ref{eq:Spectral_DM})
vanishes).  This assumes small values of $\Im m\Sigma$ near $\omega=E^D_{\mathbf{k}n}$.  
If $\Im m\Sigma$ is not small at $E^D$, the QP peak can be strongly broadened.  
The value of $E_D$ is usually shifted from $E^{RS}_{\mathbf{k}n}$ (Eq.\eqref{eq:ERS})
by a non-negligible $\omega$-dependence of $\Re e \Sigma(\mathbf{k}n,\omega)$.
An additional possible slight shift of the QP peak in Eq.(\ref{eq:Spectral_DM}) can occur 
if the $\omega$-dependence of $\Im m\Sigma$ is not negligible at $E^D$.
Eq.(\ref{eq:Spectral_DM}) usually also gives
one broad satellite mainly above the quasi-particle peak (for the conduction states),
or below it (for the valence states). 

A QP part of the spectral function, in Dyson-Migdal theory, can be separated out by Taylor
expanding $\Sigma^{\rm Fan}$ around $\omega=E^D_{\mathbf{k}n}$, keeping only
the constant term in the imaginary part and both constant and linear terms in the real part.
The answer is
\begin{equation}
A_{\rm DM}^{\rm QP}(\mathbf{k}n,\omega)=\frac{Z^D_{\mathbf{k}n}}{\pi}
\frac{ \Gamma^D_{\mathbf{k}n}}
{(\omega-E^D_{\mathbf{k}n})^2 + (\Gamma ^D_{\mathbf{k}n})^2}
\label{eq:ADMQP}
\end{equation}
where the QP energy is 
\begin{equation}
E^{D}_{\mathbf{k}n}= 
\epsilon_{\mathbf{k}n}+\Re e \Sigma(\mathbf{k}n,\omega=E^D_{\mathbf{k}n}),
\label{eq:EDM}
\end{equation}
the quasiparticle weight $Z_D$ is
\begin{equation}
Z^{D}_{\mathbf{k}n}=\left( 1- \Re e \frac{\partial \Sigma(\mathbf{k}n,\omega)}{\partial \omega} 
\bigg\rvert_{\omega=E^D_{\mathbf{k}n}} 
\right)^{-1},
\label{eq:ZDM}
\end{equation}
and the QP broadening is
\begin{equation}
\Gamma^D_{\mathbf{k}n}=Z^{D}_{\mathbf{k}n}| \Im m\Sigma(\mathbf{k}n,\omega=E^D_{\mathbf{k}n})|.
\label{eq:GammaQP}
\end{equation}
In case $T=0$ and the imaginary part of the self-energy vanishes at the QP energy,
Eqs.(\ref{eq:EDM}) and (\ref{eq:ZDM}) are the exact location and 
weight of a scaled Dirac delta peak
$Z^{D}_{\mathbf{k}n} \delta(\omega-E^{D}_{\mathbf{k}n})$.

Often\cite{Martin2016,Hedin1999}, Eq.(\ref{eq:EDM}) is linearized to find an approximate
quasiparticle energy and the related weight from quantities defined at the bare eigenenergy $\epsilon_{\mathbf{k}n}$:
\begin{equation}
E^{Dlin}_{\mathbf{k}n}=
\epsilon_{\mathbf{k}n}+ Z^{Dlin}_{\mathbf{k}n} \Re e \Sigma(\mathbf{k}n,\omega=\epsilon_{\mathbf{k}n}),
\label{eq:EDMlin}
\end{equation}
with 
\begin{equation}
Z^{Dlin}_{\mathbf{k}n}=\left( 1- \Re e \frac{\partial \Sigma(\mathbf{k}n,\omega)}{\partial \omega}
\bigg\rvert_{\omega=\epsilon_{\mathbf{k}n}}
\right)^{-1}.
\label{eq:ZDMlin}
\end{equation}

The Lehmann representation of the spectral function is derived from Eqs.(\ref{eq:A},\ref{eq:GR})
using the exact eigenstates $|m\rangle$, of energy $E_m$, of the full Hamiltonian,
\begin{eqnarray}
A({\mathbf{k}n},\omega)&=& \sum_{m',m}\frac{e^{-\beta E_m}}{Z} \left[ |\langle m'|c_{\mathbf{k}n}^+| m\rangle|^2 
\delta(\omega-E_{m'}+E_m) \right. \nonumber \\
&+& \left. |\langle m' |c_{\mathbf{k}n} | m\rangle|^2  \delta(\omega+E_{m'}-E_m) \right].
\label{eq:ALeh}
\end{eqnarray}
 where Z is the partition function, $\Sigma_m \exp(-\beta E_m)$ and $\beta$ is $1/k_BT$.
Only $T=0$ is directly relevant to the computations of this paper, but the $T>0$
properties are also important and interesting. From Eq.(19)
it is clear that $A({\mathbf{k}n,\omega})\ge 0$ at all $\omega$, and the integrated spectral weight  ($\int d\omega A({\mathbf{k}n},\omega)$) is 1.
The QP part (Eq.(\ref{eq:ADMQP})) has total weight $Z^D_{\mathbf{k}n}$ which must therefore be less than 1.
The non-QP part $A-A_{\rm QP}$ has the rest of the spectral weight.  

There is an interesting property of the first moment \cite{Mayers2016},
\begin{equation}
\int_{-\infty}^\infty d\omega \omega A({\mathbf{k}n},\omega) = \langle \{ [c_{\mathbf{k}n},H],c_{\mathbf{k}n}^+\}\rangle.
\label{eq:1moment}
\end{equation}
For noninteracting electrons, the right-hand side of Eq.(\ref{eq:1moment}) is the band energy $\epsilon_{\mathbf{k}n}$.
If the only interaction is with phonons, the right-hand side has an extra piece,
$ \langle \{ [c_{\mathbf{k}n},H_{e-p}],c_{\mathbf{k}n}^+\}\rangle$.  Terms in the electron-phonon interaction $H_{e-p}$
which have odd powers of lattice displacement do not contribute.  The even powers do,
however, and the total right hand side of Eq.(\ref{eq:1moment}) is $\langle {\mathbf{k}n}| \langle H \rangle |{\mathbf{k}n} \rangle$
where $\langle H \rangle$ is the thermal average of the vibrating one-electron Hamiltonian.  This is
exactly a Debye-Waller shifted single particle energy, both zero-point and thermal if $T>0$.  
If only the leading second-order term
in a vibrational Taylor expansion is kept, the answer is $\epsilon_{\mathbf{k}n} + \Sigma^{\rm DW}({\mathbf{k}n})$,
with only the Debye-Waller part of Eq.(\ref{self_energy_DW+Fan}).  It is interesting that an observable (in principle) 
property can separate the Debye-Waller from the Fan effects, given that translational invariance
forces a partial cancellation \cite{Allen1976} and indicates an underlying connection between these terms.

%%%%
Although the quasiparticle energy from the D-M spectral function
%\cite{note-selfconsistency} 
for the Fr\"ohlich problem
does not occur at the right place and the D-M quasiparticle spectral weight is quite bad, as shown by Figs. 1 and 2, 
the integral of the D-M spectral function is correctly 1,
and its first moment is correctly $\epsilon_{\mathbf{k}n}$ for the Fr\"ohlich problem which has no 2-phonon
matrix element and therefore no D-W term.

\section{Cumulant}\label{sec:cumulant}

\begin{figure}
\centering
\includegraphics[width=0.50\textwidth]{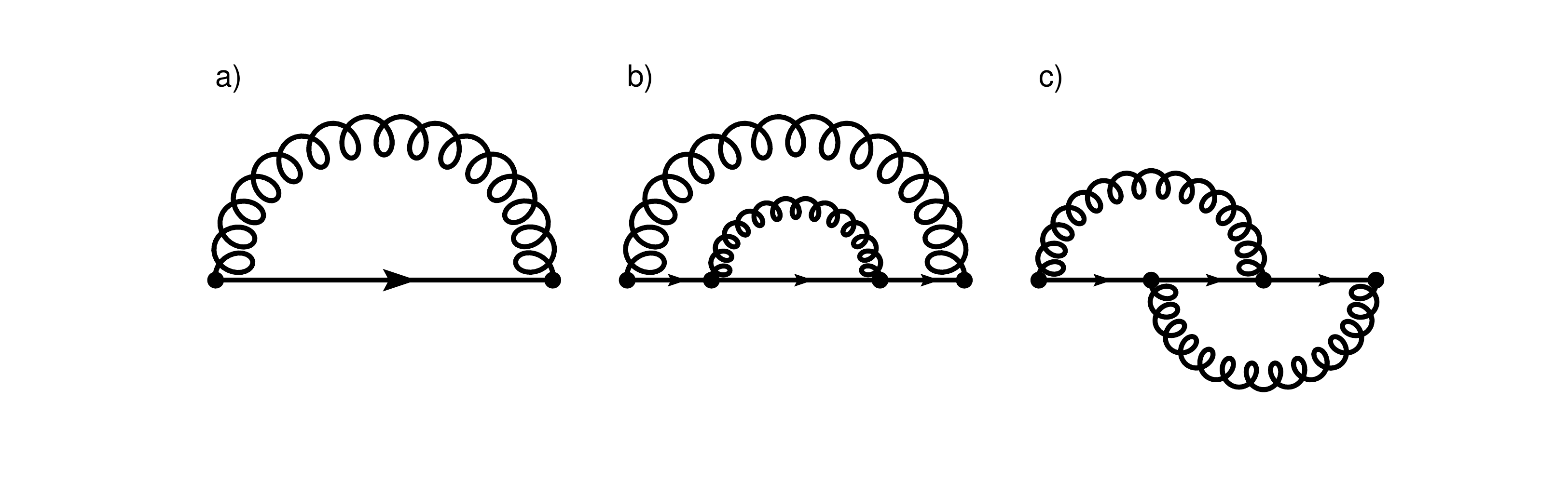}
\caption{Some of the lowest order Feynman diagrams. (a) Usual Fan self energy diagram. (b) Diagram that contributes 
to the self-consistency of the electron propagator. (c) Diagram that contains a vertex correction. \label{diagrams}}
\end{figure}

The cumulant expansion is an alternative to the usual Dyson diagrammatic perturbation theory. It 
derives from an exponential representation of the Green's function $G(\mathbf{k}n,t)$,
either retarded or t-O, in the time domain: 
\begin{eqnarray}
G_C(\mathbf{k}n,t) &=& G_0(\mathbf{k}n,t) e^{C(\mathbf{k}n,t)}.
\label{eq:G_C}
\end{eqnarray}
%{\color{red} I deleted the rest of this sentence -- PBA.}
%where $t$ represents the time difference between the annihilation and creation of the electron.
To lowest order, it treats the Fan diagram, Fig. \ref{diagrams} (a), exactly, and higher-order
diagrams such as (b) and (c), approximately.

Different methods are used to derive the cumulant $C(\mathbf{k}n,t)$ from the self energy, including
identification of the terms of the same power of the interaction in a diagrammatic expansion 
of the left- and right-hand side of Eq.(\ref{eq:G_C}).
In the case of one isolated electronic level (without dispersion), the cumulant
approach gives the exact result using only a second-order self energy \cite{Langreth1970}.
Among others, Langreth \cite{Langreth1970}, Hedin \cite{Hedin1980}, Gunnarsson {\it et al.} \cite{Gunnarsson1994},
Aryasetiawan {\it et al.} \cite{Aryasetiawan1996}, Guzzo {\it et al.} \cite{Guzzo2011} and Kas {\it et al.} \cite{Kas2014}
examined, developed and applied the cumulant expansion. 

%While prior work focused on the use of cumulant expansion of the t-O Green's function,
Prior cumulant work supposed decoupling between empty and occupied states, and worked in a t-O formulation.
In a recent paper, Kas {\it et al.} \cite{Kas2014} considered
a cumulant expansion of the retarded Green's function.
In the case of metals, they find that the retarded version correctly includes recoil effects
that mix particle and hole states, while these do not appear within the existing cumulant 
approaches based on t-O Green's functions where particle and hole states are treated separately.
Their equations, applied to the EPI self energy (including Fan and Debye-Waller contributions), are
\begin{eqnarray}
&&A^R_C(\mathbf{k}n,\omega) = -\frac{1}{\pi} {\Im m} G^R_C(\mathbf{k}n,\omega),
\label{eq:ARC_omega}
\\
&&G^R_C(\mathbf{k}n,\omega) = \int_{-\infty}^\infty e^{i \omega t} G^R_C(\mathbf{k}n,t) dt,
\label{eq:GRC_omega}
\\
&&G^R_C(\mathbf{k}n,t) = -i \theta (t) 
e^{-i(\varepsilon_{\mathbf{k}n} + \Sigma^{\mathrm{DW}}_{\mathbf{k}n})t}e^{C^R(\mathbf{k}n,t)},
\label{eq:GRC_t}
\\
&&C^R(\mathbf{k}n,t) = \int_{-\infty}^\infty \beta^R(\mathbf{k}n,\omega) \frac{e^{-i \omega t} + i \omega t - 1}{\omega^2} d\omega,
\label{eq:CR_t}
\\
&&\beta^R(\mathbf{k}n,\omega) = \frac{1}{\pi} 
 | \Im m \, \Sigma^{\mathrm{Fan}}(\mathbf{k}n,\omega +\varepsilon_{\mathbf{k}n})|.
\label{eq:BR_omega}
\end{eqnarray}
Eq.(\ref{eq:GRC_t}) shows that
the static Debye-Waller self energy $\Sigma^{\mathrm{DW}}_{\mathbf{k}n}$ shifts (in frequency) 
the whole spectral function with respect to $\varepsilon_{\mathbf{k}n}$.  
The retarded character of the Green's function is present in Eq.(\ref{eq:BR_omega}), where both
unoccupied and occupied states contribute (compare with Eqs.\eqref{eq:BTOC96_omega_e} 
and \eqref{eq:BTOC96_omega_h}, given later). 
Another effect of the retarded character of this cumulant 
is the limits of frequency integrals in Eq.(\ref{eq:CR_t}) (compare with 
Eqs.\eqref{eq:CTOC96_t_e} and \eqref{eq:CTOC96_t_h}).

The evaluation of Eq.(\ref{eq:CR_t}) might appear delicate due to possible numerical problems associated with the
square of the frequency in the denominator. However, the numerator and its first derivative also vanish for $\omega=0$.
The same Eq.(\ref{eq:CR_t}) also shows that the cumulant and its first time derivative vanish at $t=0$,
from which one deduces that the integral of the spectral function $A^R_C(\mathbf{k}n,\omega)$ over all frequencies 
is 1, 
%normalized, 
while its first moment is equal to $\varepsilon_{\mathbf{k}n} + \Sigma^{\mathrm{DW}}_{\mathbf{k}n}$.
%The dynamical part of the self energy does not change this first moment.
These properties agree perfectly with the exact results of the previous section.

Moreover, for the EPI, the phonon density of states (DOS) 
from acoustic modes vanishes quadratically at zero frequency, and translation invariance
shows that these modes have non-diverging electron-phonon coupling.
Thus, for the top of the valence band and the bottom of the conduction band,
$\beta^R(\mathbf{k}n,\omega)$ vanishes quadratically around $\omega=0$.
Then following the method of Ref.~\onlinecite{Aryasetiawan1996}, three separate
effects can be attributed to the three pieces ($e^{-i\omega t}+i\omega t -1$) of Eq.(\ref{eq:CR_t}).
Specifically, $e^{-i\omega t}$ generates the satellites, $+i\omega t$ shifts the quasiparticle peak,
and $-1$ generates the quasiparticle weight.  The latter two effects make use of Kramers-Kronig relations,
\begin{eqnarray}
&&\Re e \Sigma^{\mathrm{Fan}}(\mathbf{k}n,\varepsilon_{\mathbf{k}n})
= - P \int_{-\infty}^\infty \frac{\beta^R(\mathbf{k}n,\omega)}{\omega} d\omega,
\label{eq:ReSigma}
\\
&&\Re e \frac{\partial \Sigma^{\mathrm{Fan}}(\mathbf{k}n,\omega)} {\partial \omega} |_{\omega=\varepsilon_{\mathbf{k}n}} 
= - \Re e \int_{-\infty}^\infty  \frac{\beta^R(\mathbf{k}n,\omega)}{(\omega+i \delta)^2} d\omega,
\nonumber
\\ 
\label{eq:DerivSigma}
\end{eqnarray}
%
%These quantities directly give access to 
where $P$ denotes the principal part of the integral.
The first of these gives the quasiparticle peak shift,
\begin{eqnarray}
E^{R}_{\mathbf{k}n}&=& \epsilon_{\mathbf{k}n}
+\Re e \Sigma^{\mathrm{Fan}}(\mathbf{k}n,\omega=\epsilon_{\mathbf{k}n})+\Re e \Sigma^{\mathrm{DW}}(\mathbf{k}n)
\nonumber \\
&=& \epsilon_{\mathbf{k}n} +\Re e \Sigma(\mathbf{k}n,\omega=\epsilon_{\mathbf{k}n}).
\label{eq:ER}
\end{eqnarray}
This is identical to the shift $E^{RS}_{\mathbf{k}n}$ from Rayleigh-Schr\"odinger
perturbation theory.  The second contributes the quasiparticle weight,
\begin{equation}
Z^{R}_{\mathbf{k}n}= 
\exp\bigg(\Re e \frac{\partial \Sigma^{\mathrm{Fan}}(\mathbf{k}n,\omega)} 
{\partial \omega} |_{\omega=\varepsilon_{\mathbf{k}n}}\bigg).
\label{eq:ZR}
\end{equation}
%
%Finally, the imaginary exponential in Eq.(\ref{eq:CR_t}) yields the satellites.

Thanks to Eqs.(\ref{eq:ReSigma}-\ref{eq:DerivSigma}), Eq.(\ref{eq:CR_t}) can be rewritten as
\begin{eqnarray}
C^{R}(\mathbf{k}n,t) &=& 
 \int_{-\infty}^\infty \beta^{R}(\mathbf{k}n,\omega) 
e^{-i \omega t}\Re e\frac{1}{(\omega+i\delta)^2} d\omega
\nonumber \\
&-&it\Sigma^{\mathrm{Fan}}(\mathbf{k}n,\varepsilon_{\mathbf{k}n}) 
+ \frac{\partial \Sigma^{\mathrm{Fan}}(\mathbf{k}n,\omega)} {\partial \omega} |_{\omega=\varepsilon_{\mathbf{k}n}}.
\nonumber \\
\label{eq:CR_t2}
\end{eqnarray}

In the t-O cumulant approach of Ref.~\onlinecite{Aryasetiawan1996}, Eqs.(\ref{eq:CR_t2}) and (\ref{eq:BR_omega})
change as follows for the electrons:
\begin{eqnarray}
C^{t-O}_e(\mathbf{k}n,t) &=& 
 \int_{\mu-\varepsilon_{\mathbf{k}n}}^\infty \beta^{t-O}_e(\mathbf{k}n,\omega) 
e^{-i \omega t} \Re e\frac{1}{(\omega+i \delta)^2} d\omega
\nonumber \\
&-&it\Sigma^{\mathrm{Fan}}(\mathbf{k}n,\varepsilon_{\mathbf{k}n}) 
+ \frac{\partial \Sigma^{\mathrm{Fan}}(\mathbf{k}n,\omega)} {\partial \omega} |_{\omega=\varepsilon_{\mathbf{k}n}},
\nonumber \\
\label{eq:CTOC96_t_e}
\\
\beta^{t-O}_e(\mathbf{k}n,\omega) &=& \frac{1}{\pi}
 |\Im m \, \Sigma^{\mathrm{Fan}}_{un}(\mathbf{k}n,\omega +\varepsilon_{\mathbf{k}n})|.
\label{eq:BTOC96_omega_e}
\end{eqnarray}
Note the reduced range of the integral, as well as the selection of part of the self energy in the $\beta$ factor.
By contrast, the contributions that are either constant in time or linear in time are computed from the whole self energy.
Unlike the retarded cumulant, this version of the t-O cumulant does not vanish at $t=0$, nor does
its time-derivative, which means
that the spectral function is not normalized to 1, and its first moment is changed by the dynamical contribution.
Corresponding expressions for the holes are also presented in Ref.~\onlinecite{ZhouThesis2015}:
\begin{eqnarray}
C^{t-O}_h(\mathbf{k}n,t) &=& 
 \int_{-\infty}^{\mu-\varepsilon_{\mathbf{k}n}} \beta^{t-O}_h(\mathbf{k}n,\omega) 
e^{-i \omega t}\Re e\frac{1}{(\omega+i \delta)^2} d\omega
\nonumber \\
&-&it \Sigma^{\mathrm{Fan}}(\mathbf{k}n,\varepsilon_{\mathbf{k}n}) 
+ \frac{\partial \Sigma^{\mathrm{Fan}}(\mathbf{k}n,\omega)} {\partial \omega} |_{\omega=\varepsilon_{\mathbf{k}n}},
\nonumber \\
\label{eq:CTOC96_t_h}
\\
\beta^{t-O}_h(\mathbf{k}n,\omega) &=& \frac{1}{\pi}
 |\Im m \, \Sigma^{\mathrm{Fan}}_{oc}(\mathbf{k}n,\omega +\varepsilon_{\mathbf{k}n})|.
\label{eq:BTOC96_omega_h}
\end{eqnarray}

Slightly different versions of the t-O cumulant approach arise from the replacement
of the full Fan self energy and its derivative, 
in the last two terms of Eqs.(\ref{eq:CTOC96_t_e}) and (\ref{eq:CTOC96_t_h}), 
by their unoccupied and occupied counterparts, respectively, as in Ref.~\onlinecite{Gumhalter2016} 
(see Eq.36),
or the replacement of its derivative only, as in Refs.~\onlinecite{Guzzo2011,ZhouThesis2015}
(see Eqs. 3.32 and 3.33 of Ref.~\onlinecite{ZhouThesis2015}).
In both cases, the normalization is brought back.
%%%%
%\stkout{As we will see later, for our insulators, with a gap much larger than the maximum phonon frequency, 
%the difference between these t-O versions and with the retarded cumulant version is tiny.}
%%%%
 Our computations for MgO and LiF have such large band gaps that the differing limits of integration
in Eqs. \ref{eq:CR_t2}, \ref{eq:CTOC96_t_e}, and \ref{eq:CTOC96_t_h} have negligible consequences.  Similarly,
the function $\beta^R(\mathbf{k},n=c)$ is almost identical to $\beta^{t-O}_e$  and 
$\beta^R(\mathbf{k},n=v)$ is almost identical to $\beta^{t-O}_h$, in the relevant range of $\omega$-integration
where the denominator $\omega^2$ is small.  The
different versions of cumulant methods are thus sufficiently similar that only retarded cumulant results are presented here. 
%%%%

In Eq.(\ref{eq:G_C}), or equivalently Eq.(\ref{eq:GRC_t}), the exponential can be Taylor-expanded: 
the Green's function and spectral function in the time domain are the sum of the product of the 
independent-electron contribution multiplied by powers of the cumulant, the latter being a linear 
functional of the imaginary part of the self energy. In the frequency domain, this creates a satellite 
series \cite{Aryasetiawan1996, Gumhalter2016}, coming from repeated convolution 
 in frequency of the undressed 
particle spectral function (a Dirac delta) by the 
%%%%
%\stkout{cumulant function $\beta$ in the frequency domain.} 
%%%%
%{\color{red} ``satellite spectral function'' $A^S$ (Eqs. 3 and 6 of Ref. \onlinecite{Verdi2017}).
%Supposing $A^S$ to be also a Dirac delta,}
%%%
  ``satellite spectral function'' $A^S$ 
(See Eqs. 8, 13, and 14 of Ref. \onlinecite{Aryasetiawan1996} or Eqs. 3, 4, and 6 of the supplemental materials 
of Ref. \onlinecite{Verdi2017}).  Supposing $A^S$ to be also a Dirac delta (as in the Lundqvist model Hamiltonian), 
one obtains for the spectral function a Poisson distribution of Dirac function satellites, each being spaced 
by the characteristic bosonic satellite energy.
%%%%
%\stkout{Supposing the 
%imaginary part of the self-energy to be a single boson peak (also a Dirac delta), and likewise the $\beta$ function,}
%%%%
%one obtains for the spectral function a Poisson distribution of Dirac function satellites \cite{Langreth1970}, 
%each being spaced by the characteristic {\color{red}Bosonic satellite} energy. 
More generally, supposing the imaginary 
part of the self-energy to be non-zero only for positive frequencies beyond a given threshold, 
the threshold for each satellite contribution will be determined by the self-energy threshold multiplied by the order of the satellite. 
In the next section, one will see such an effect in the case of the Fr\"ohlich Hamiltonian. 

When the imaginary part of the self energy has contributions at both negative and positive frequencies, this simple picture is lost, unless the cumulant is clearly dominated by one of these. In the case of the first-principles EPI for wide-gap insulators, the self energy has indeed both negative and positive contributions. However, only one of these contributions will dominate for wide-gap insulators, as will be seen later.

\section{FR\"OHLICH HAMILTONIAN SELF-ENERGY}

The relationship between the Fr\"ohlich coupling and first-principles calculations has been 
established in Refs.~\onlinecite{Sjakste2015,Verdi2015}, in the general case of several electronic bands, several
phonon branches, as well as anisotropic Born effective charge tensor and dielectric tensor.
Here, we consider the simple original Fr\"ohlich Hamiltonian, corresponding to the following hypotheses:
(1) Only one isotropic electronic parabolic band (we will first treat the
conduction band), with a minimum at $\mathbf{k}=0$.
We choose $\varepsilon_{\mathbf{k}=0,c}=0$, and use parabolic dispersion governed by
the effective mass $m^*$. 
(2) Only one LO-phonon branch with constant phonon frequency $\omega_{LO}$.
(3) Isotropic Born effective charge $Z^*$, isotropic electronic (optical) dielectric constant $\epsilon_\infty$,
and thus, isotropic low-frequency dielectric constant $\epsilon_0$ \cite{Gonze1997}.

Only intraband terms $n^\prime =n$ are present, thus, the general
$g_{nn'j}(\mathbf{k},\mathbf{q})=\langle \mathbf{k}n|H^{(1)}_{\mathbf{q}j}|\mathbf{k+q}n'\rangle$ reduces to
$g_{nn'j}(\mathbf{k},\mathbf{q})=g_{\mathbf{q}} \delta_{nn'}$ for $j=LO$, with
\begin{equation}
g_{\mathbf{q}}= \frac{i}{q}\Bigg[\frac{4 \pi}{\Omega_0}  \frac{\omega_{LO}}{2} 
\Big(  \frac{1}{\epsilon_\infty}-\frac{1}{\epsilon_0}   \Big)    \Bigg]^{1/2},
\label{eq:FrohlichEPI}
\end{equation}
where $\Omega_0$ is the volume of the primitive cell. 

In computing the self energy, the constant Debye-Waller shift 
is neglected. This might seem a strong approximation. However, hypothesis (2) 
implies the neglect of the Fan term from the acoustic modes as well, and it is known that 
the acoustic mode Fan contributions and the Debye-Waller contributions cancel 
each other in the vanishing-q limit \cite{Allen1976}. Furthermore, the LO-phonon Fan term dominates, due to the integrable divergence mentioned at the end of Sec. II.

The expression for the Fr\"ohlich 
self energy is easily found, see {\it e.g.} Ref.~\onlinecite{Mahan2000}, 
but is presented here as well, for convenience and comparison with the
first-principle results. 
The zero-temperature formula for the Fan self energy of the (Fr\"ohlich) electron state at the bottom of the conduction
band ($\mathbf{k}=0$) comes from Eq.(\ref{self_energy_Fan_un}):
\begin{equation}
\Sigma_{F,e}(\mathbf{k}=0,\omega)=\frac{1}{N_{\mathbf{q}}}\sum_{\mathbf{q}} \frac{|g_{\mathbf{q}}|^2}
{\omega- \epsilon_{\mathbf{q}} -\omega_{LO} +i\eta }.
\label{eq:G}
\end{equation}
The intermediate electron energy $\epsilon_{\mathbf{q}}$ becomes $q^2/2m^\ast$ in the parabolic band with 
effective mass approximation.  
Using the Debye sphere for the Brillouin zone, the equation becomes
\begin{equation}
\Sigma_{F,e}(0,\omega)=\int_0^{q_D}
dq \frac{\Omega_0}{(2 \pi)^3}\frac{4 \pi q^2 |g_q|^2}{\omega-\frac{q^2}{2m^\ast}-\omega_{LO}+i\eta}.
\label{eq:G1}
\end{equation}
Provided the electronic energy on the boundary of the Debye sphere is much bigger
than $\omega_{LO}$, the upper limit $q_D$ can be safely extended to infinity \cite{Mahan2000}.
For $q_D \rightarrow \infty$, this gives
\begin{eqnarray}
\Re e \, \Sigma_{F,e}(\omega)&=&-\frac{\alpha \omega_{LO}}{\sqrt{1-\omega/\omega_{LO}}}\theta(\omega_{LO}-\omega),
\label{eq:ReSigmaFrohlich_e}
\\
\Im m \, \Sigma_{F,e}(\omega)&=&-\frac{\alpha \omega_{LO}}{\sqrt{\omega/\omega_{LO}-1}}\theta(\omega-\omega_{LO}),
\label{eq:ImSigmaFrohlich_e}
\end{eqnarray}
where the Fr\"ohlich coupling constant $\alpha$ is
\begin{equation}
\alpha= 
\Big( \frac{1}{\epsilon_\infty}-\frac{1}{\epsilon_0} \Big)
\left(\frac{m^\ast}{2\omega_{LO}}\right)^{1/2}.
\label{eq:alpha}
\end{equation}
Above $\omega_{LO}$, the self energy is purely imaginary, while below $\omega_{LO}$,
it is purely real. Both are negative and diverge like an inverse square root of the frequency around 
$\omega_{LO}$.

%If, instead of treating a conduction band in hypothesis (1), we treat a 
For the valence band, with
the eigenenergy of the top of the valence band now taken as zero, 
%we can compute straightforwardly its 
the corresponding retarded self energy is
%with only hole contributions,
%
\begin{eqnarray}
\Re e \, \Sigma_{F,h}(\omega)&=&+\frac{\alpha \omega_{LO}}{\sqrt{1+\omega/\omega_{LO}}}\theta(\omega+\omega_{LO}), 
\label{eq:ReSigmaFrohlich_h}
\\
\Im m \, \Sigma_{F,h}(\omega)&=&-\frac{\alpha \omega_{LO}}{\sqrt{-\omega/\omega_{LO}-1}}\theta(-\omega-\omega_{LO}).
\label{eq:ImSigmaFrohlich_h}
\end{eqnarray}
In the t-O case, the hole self energy has imaginary part of opposite sign. 
For a given material with well-defined dielectric constant and LO frequency, 
the coupling constant $\alpha$ from Eq. \ref{eq:alpha}
has different values for electron and hole polarons, due to differing effective masses.

\begin{figure}
\centering
\includegraphics[width=0.50\textwidth]{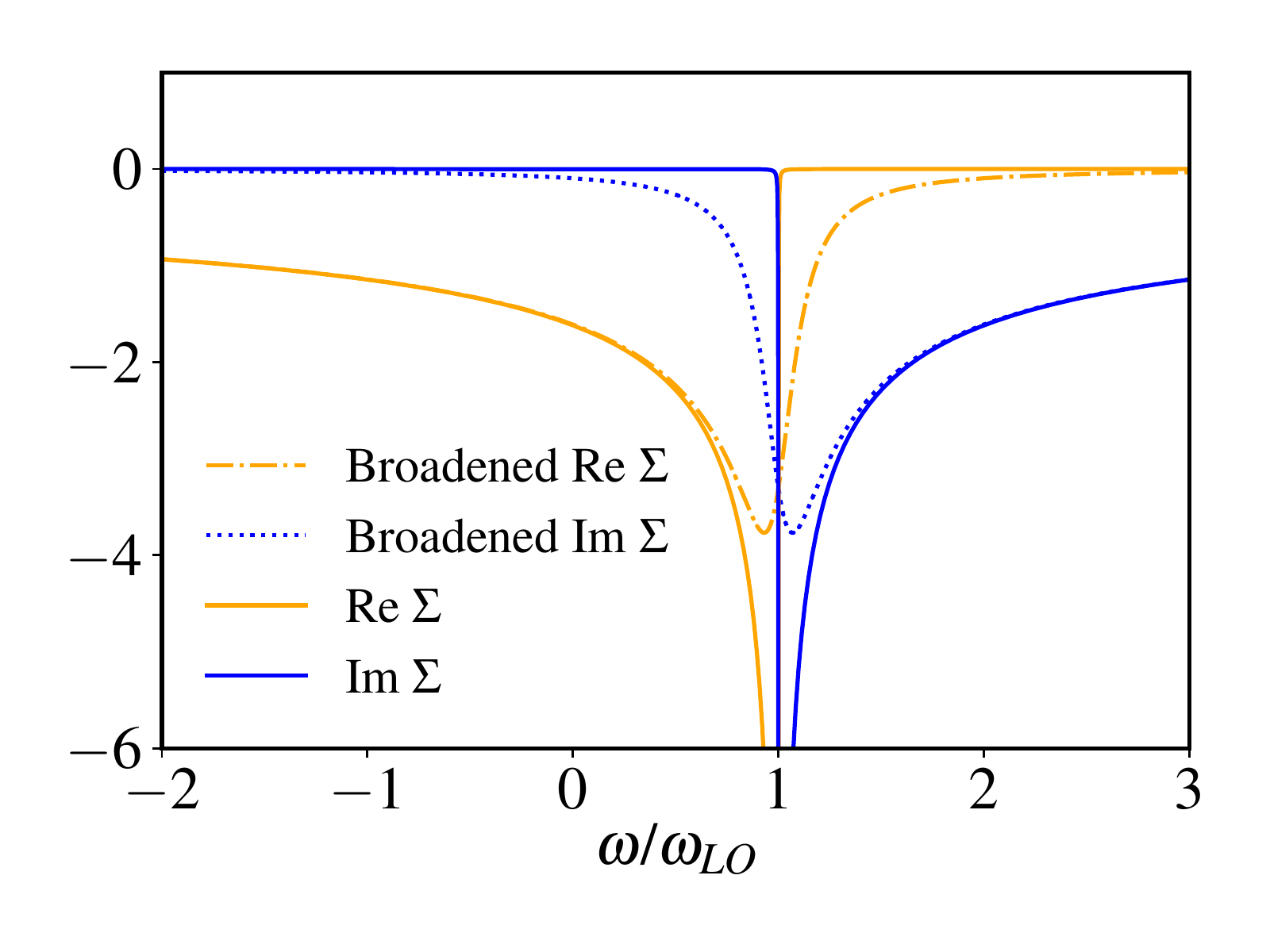}
\caption{Fr\"ohlich Hamiltonian self-energy. Real part in orange, imaginary part in blue. The functions with a negligible broadening, $\eta=0.001\omega_{LO}$,
are represented by continuous lines, while functions with a broadening $\eta=0.12\omega_{LO}$, similar to the
one used in first-principles calculations, are represented by dashed lines.
\label{fig:SpectralFrohlich}}
\end{figure}

In Secs. VII and VIII dealing with first-principles calculations, we will maintain a small finite broadening factor
$\eta$, of order 0.12$\omega_{LO}$, for numerical reasons.          
Thus the self-energy functions, Eqs.(\ref{eq:ReSigmaFrohlich_e}) and (\ref{eq:ImSigmaFrohlich_h}), will 
not retain their inverse-square-root shape. Eqs.(\ref{eq:ReSigmaFrohlich_e}) and (\ref{eq:ImSigmaFrohlich_e}) and their broadened versions
are represented in Fig.~\ref{fig:SpectralFrohlich}.

The self energies in 
Eqs.(\ref{eq:ReSigmaFrohlich_e})-(\ref{eq:ImSigmaFrohlich_h})
were derived in lowest order perturbation theory.
They include only the Fan diagram, without vertex corrections. Calculations of self energies at higher orders 
have been performed for the Fr\"ohlich Hamiltonian,
see {\it e.g.} Refs.~\onlinecite{Mahan2000,Dunn1975}.  
For values of $\alpha$ in the range considered in the present paper,
those higher-order corrections to the \stkout{D-M version of the} quasiparticle shift are small, 
consistent with Migdal \cite{Migdal1958}.
Calculations using these formulas for $\Sigma_F$ and the corresponding spectral functions 
$A_{\rm DM}$ and $A_{\rm C}$, are discussed in the next section, and plotted in 
Figs. \ref{spectral_functions_alpha0.34}-\ref{spectral_functions_alpha8}.

\section{FR\"OHLICH HAMILTONIAN QUASIPARTICLE ENERGY AND SPECTRAL FUNCTION}

\begin{figure}
\centering
\includegraphics[width=0.45\textwidth]{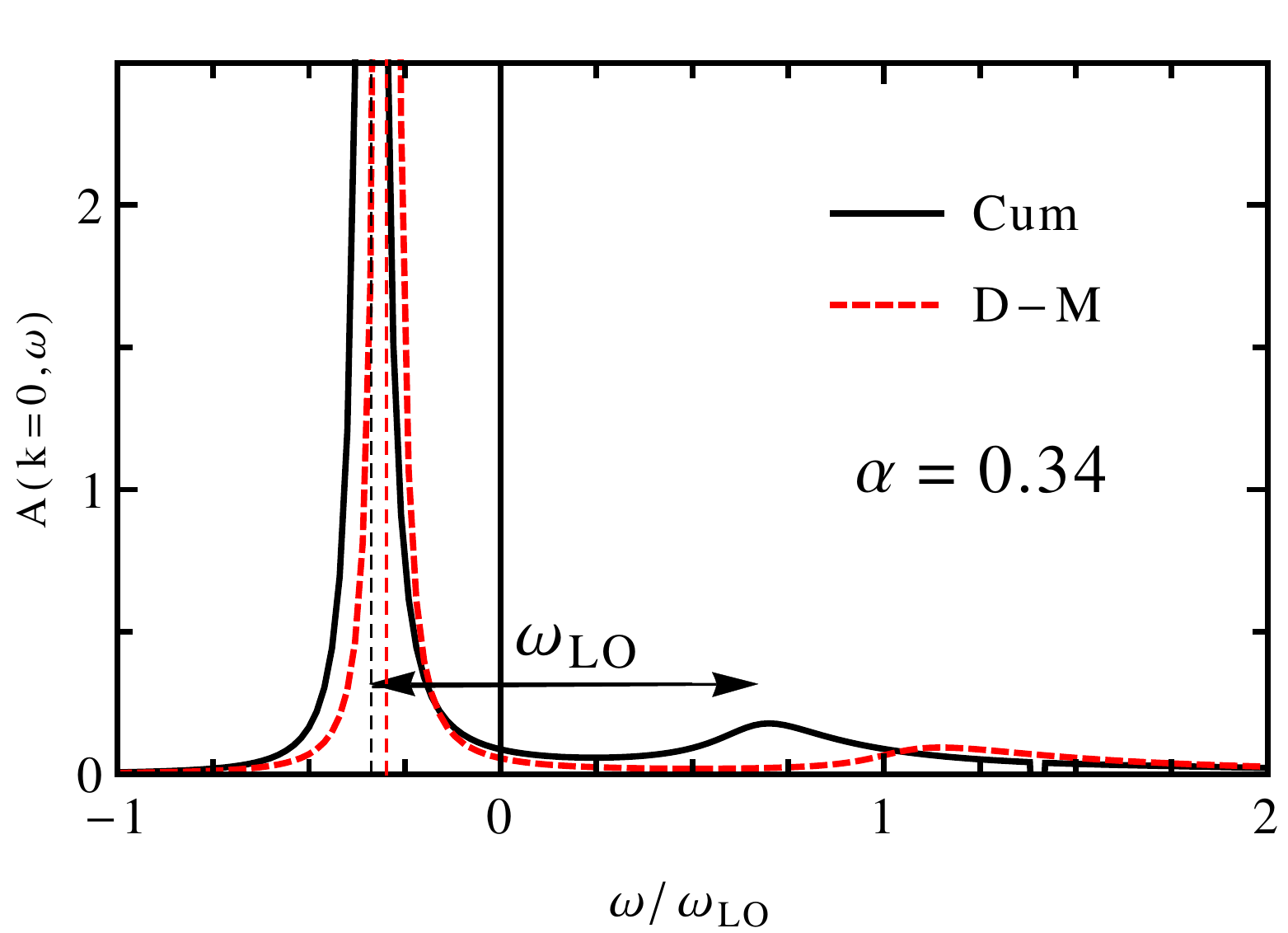}
\caption{
Fr\"ohlich Hamiltonian spectral function using the cumulant approach $G_C$ (solid, black) and the Dyson-Migdal 
approach $G_D$ (dashed, red) for $\alpha$=0.34.
The position of the quasiparticle peak slightly differs between the two.  The cumulant
version deviates from the Fr\"ohlich value $-\alpha\omega_{\rm LO}$ only
because a non-zero broadening $\delta\approx 0.12\omega_{LO}$ \cite{note-broadening} 
is used in numerical evaluation of Eq.\eqref{eq:G1}, for consistency with later calculations.  
In the D-M case, the onset of the phonon-emission 
``satellite" is higher by $\omega_{\rm LO}$ than
the bare band energy $\omega=\varepsilon_{\mathbf{k}=0,c}^{(0)}=0$ \cite{note-selfconsistency}. 
By contrast, it is higher by $\omega_{\rm LO}$ than the quasiparticle peak  in the cumulant method,  
corresponding to states that combine the dressed quasiparticle with one LO phonon.
\label{spectral_functions_alpha0.34}}
\end{figure}

The Rayleigh-Schr\"odinger approximation Eq.(\ref{eq:ERS}) with the self energy Eq.(\ref{eq:ReSigmaFrohlich_e}),
gives the quasiparticle peak at $E^{RS}_{QP}=-\alpha\omega_{\rm LO}$
for the Fr\"ohlich Hamiltonian, at $\mathbf{k}=0$ with $\epsilon_{\mathbf{k}=0,c}=0$.
This is the standard Fr\"ohlich result \cite{Mahan2000}.
Making fuller use of the self energy Eqs.(\ref{eq:ReSigmaFrohlich_e}-\ref{eq:ImSigmaFrohlich_e}), 
the spectral function $-{\Im m}G_R(\mathbf{k}n,\omega)/\pi$ has
``dynamical effects" beyond Rayleigh-Schr\"odinger.
In Fig. \ref{spectral_functions_alpha0.34} we show the spectral function for $\alpha=0.34$ (a small value,
 typical of many semiconductors, e.g. electrons at the conduction band minimum of GaN), 
 for the Fr\"ohlich Hamiltonian using two approximations. The black curve corresponds
to the cumulant expansion Eqs.(\ref{eq:ARC_omega}-\ref{eq:BR_omega}), while the dash-red curve 
was obtained with the D-M approach Eq.(\ref{eq:Spectral_DM}).
For consistency with later first-principles calculations, we used a small broadening factor $\delta\approx 0.12\omega_{LO}$ \cite{note-broadening} for the self energy, which causes a small artificial shift and broadening of the quasiparticle peak.

We see two effects. (i) In the cumulant case, the quasiparticle peak agrees with the value $-\alpha\omega_{\rm LO}$
predicted by Rayleigh-Schr\"odinger perturbation theory.
(ii) A phonon-emission side band appears, with one clearly visible satellite, at different 
energies in the D-M and cumulant cases.
The separation between the quasiparticle and satellite is about $\omega_{LO}$ in the cumulant expansion, 
but slightly higher than $(1+\alpha) \omega_{LO}$ in the D-M approach.
Physically, this satellite ought to start at the quasiparticle energy $E_{\rm QP}$ plus $\omega_{LO}$.
The D-M shift of $E_{\rm QP}$ away from $-\alpha\omega_{\rm LO}$ does not agree with the results
obtained for many years by the polaron community \cite{Mahan2000,Devreese2009}.

Increasing the value of $\alpha$ to values typical for the valence and conduction band extrema of MgO and LiF,
 1.62, 4.01 and 8.00 (see later) gives the three next figures
\ref{spectral_functions_alpha1},
\ref{spectral_functions_alpha4} and \ref{spectral_functions_alpha8}.
Note that the position $E_{\rm QP}$ of the quasiparticle peak 
differs more and more between the two, with the cumulant
version staying at the RS answer $-\alpha\omega_{\rm LO}$, as expected. 
$A_{\rm DM}$ becomes increasingly unphysical for stronger couplings:
the side band has one broad satellite,
setting in at $\omega_{\rm LO}$, with a maximum at a frequency that increases with $\alpha$.
By contrast, $A_C$ has the satellite onset at $E_{\rm QP}+\hbar\omega_{\rm LO}$.
Several satellites are clearly visible in $A_{\rm C}$ of Fig.~\ref{spectral_functions_alpha4}, spaced approximately by $\omega_{\rm LO}$, 
a physically sensible behavior. The side bands
become broader and less well-defined as $\alpha$ increases, with a long tail extending to higher energies.
The numerical value of the broadening $\delta\approx 0.12\omega_{LO}$, albeit small, has an impact on the threshold at which it becomes impossible to distinguish 
the satellites from the overall smooth behaviour. Indeed, this broadening factor is multiplied by the
order of the satellite in the repeated convolution of the undressed particle spectral function
mentioned in Sec. IV.

These cumulant results globally agree with the previous cumulant-based study by Dunn \cite{Dunn1975}, for $\alpha=2,4$, and $6$. 
He worked, however, at finite temperature, and also included 
the next order of perturbation theory in his calculations of the self energy.  Higher orders 
of perturbation theory tend to sharpen features of the spectral function.   
In the case of a model core-electron spectrum, for which the exact solution is known~\cite{Langreth1970}, 
the next order of perturbation theory improves significantly the position of the peak, 
and sharpens it with respect to a first-order self-consistent treatment.
A first-order non-self-consistent treatment also gives a sharper  plasmon satellite
than the first-order self-consistent treatment, albeit
located at nearly the same too low energy~\cite{Hedin1991}.

The first satellite shape and position, in our cumulant calculations,
resemble reasonably well those of the
diagrammatic Monte Carlo (MC) calculations of 
Mischenko {\it et al.}\cite{Mishchenko2000}, apparently the best reference results available at present.
However, the MC results do not show the
second and third satellites and instead develop a satellite in the range from 3.5 to 4.0 
$\omega_{LO}$ if $\alpha$ is larger than one. For values of $\alpha$ larger than 4, 
another satellite appears in the range 8.0 ... 9.0 $\omega_{LO}$.
As MC results for optical $\sigma(\omega)$ compare well with other
approaches \cite{Klimin2016}, we believe that the cumulant approach for $A(k,\omega)$
has some errors for $\omega$ beyond the first satellite.  
The physical reason for the disappearance of the second and third multiphonon peaks, 
and the appearance of other peaks, has been discussed in Ref.~\onlinecite{Mishchenko2003}. The
new peaks relate to so-called ``relaxed excited states'', not treated by the lowest order cumulant approach, 
that dominate the spectral function in the energy range beyond the first phonon threshhold.

\begin{figure}
\centering
\includegraphics[width=0.45\textwidth]{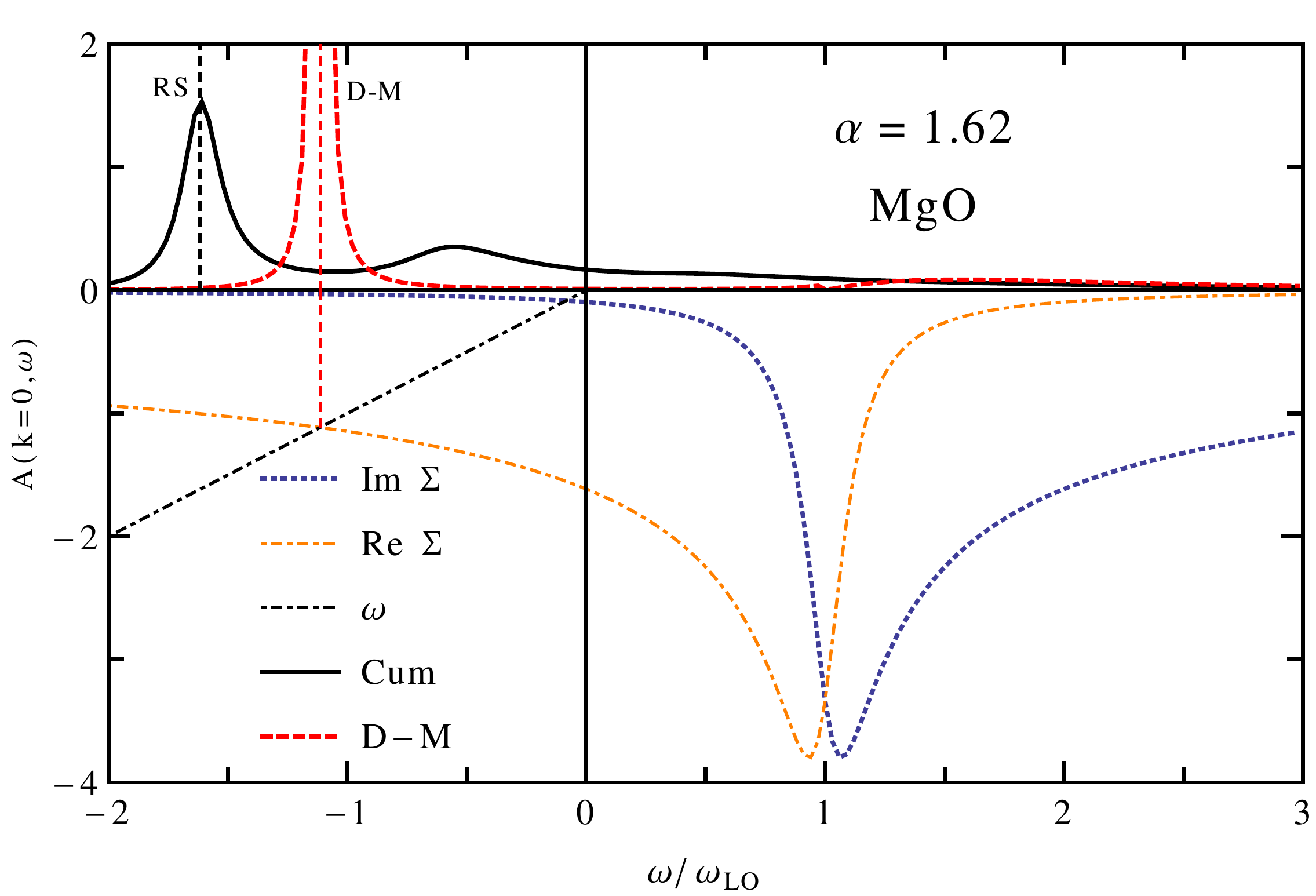}
\caption{The lower part shows the D-M self energy (in units of $\omega_{LO}$) for a Fr\"ohlich electron with
$\alpha=1.62$ (typical of the conduction band minimum of MgO), using Eq.\eqref{eq:ReSigmaFrohlich_e} except broadened with $\delta=0.12\omega_{LO}$ in
Eq.\eqref{eq:G}.  The position of the
D-M QP peak is at the crossing between the real part and the line $\Re e \Sigma=\omega$.  The upper part shows both the
resulting D-M spectral function and the cumulant version from $G_C$.  The satellite setting in at $\omega/\omega_{LO}=1$
in the D-M case is barely visible in this picture. 
\label{spectral_functions_alpha1}
}
\end{figure}
\begin{figure}
\centering
\includegraphics[width=0.45\textwidth]{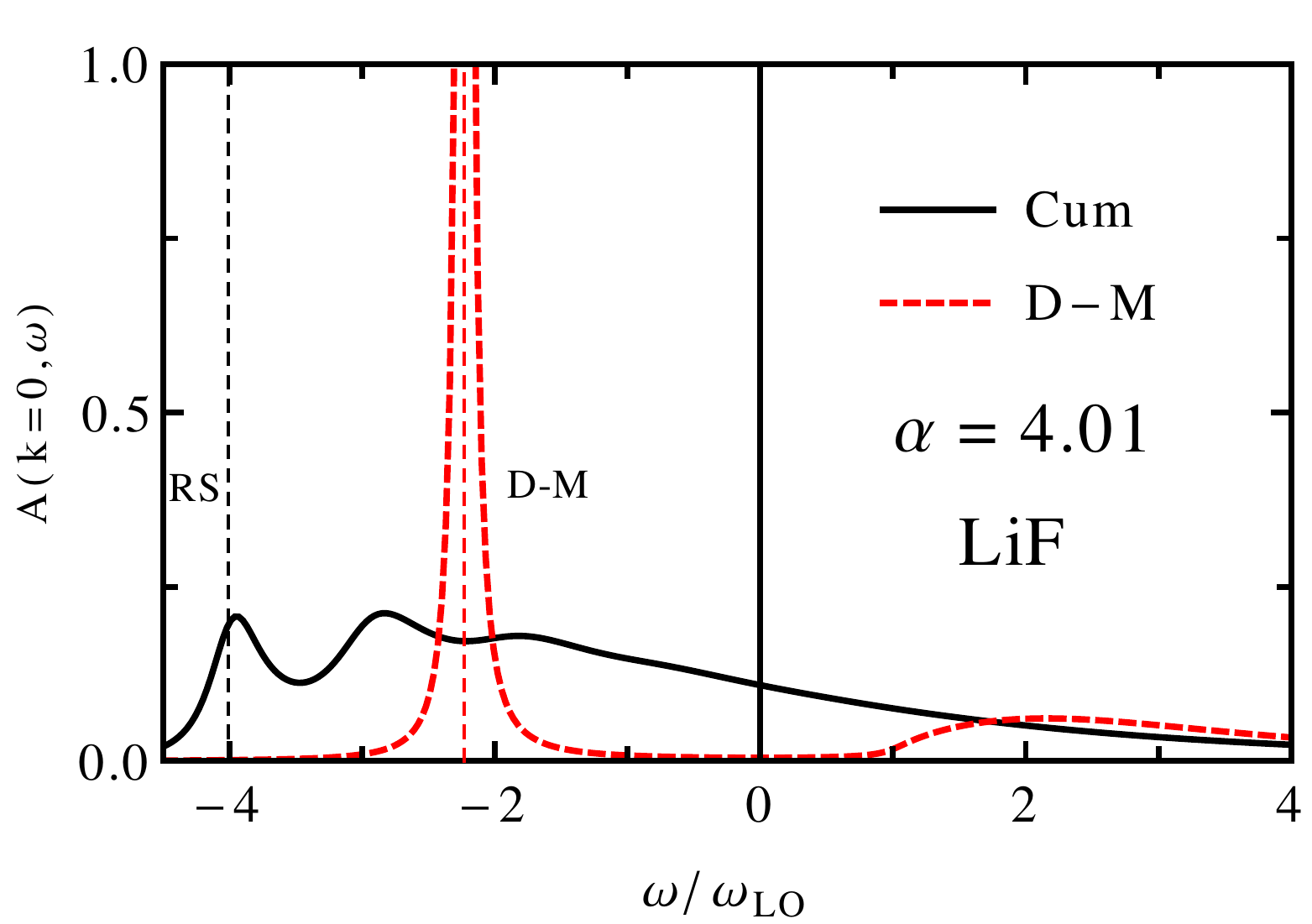}
\caption{Fr\"ohlich Hamiltonian spectral function using the cumulant $G_C$ (in black) and the 
Dyson-Migdal approach $G_D$ (dashed, in red) for $\alpha$=4.01 (typical of the conduction band minimum 
of LiF), from the Migdal self energy broadened by $\delta=0.12\omega_{LO}$.
\label{spectral_functions_alpha4}}
\end{figure}
\begin{figure}
\centering
\includegraphics[width=0.45\textwidth]{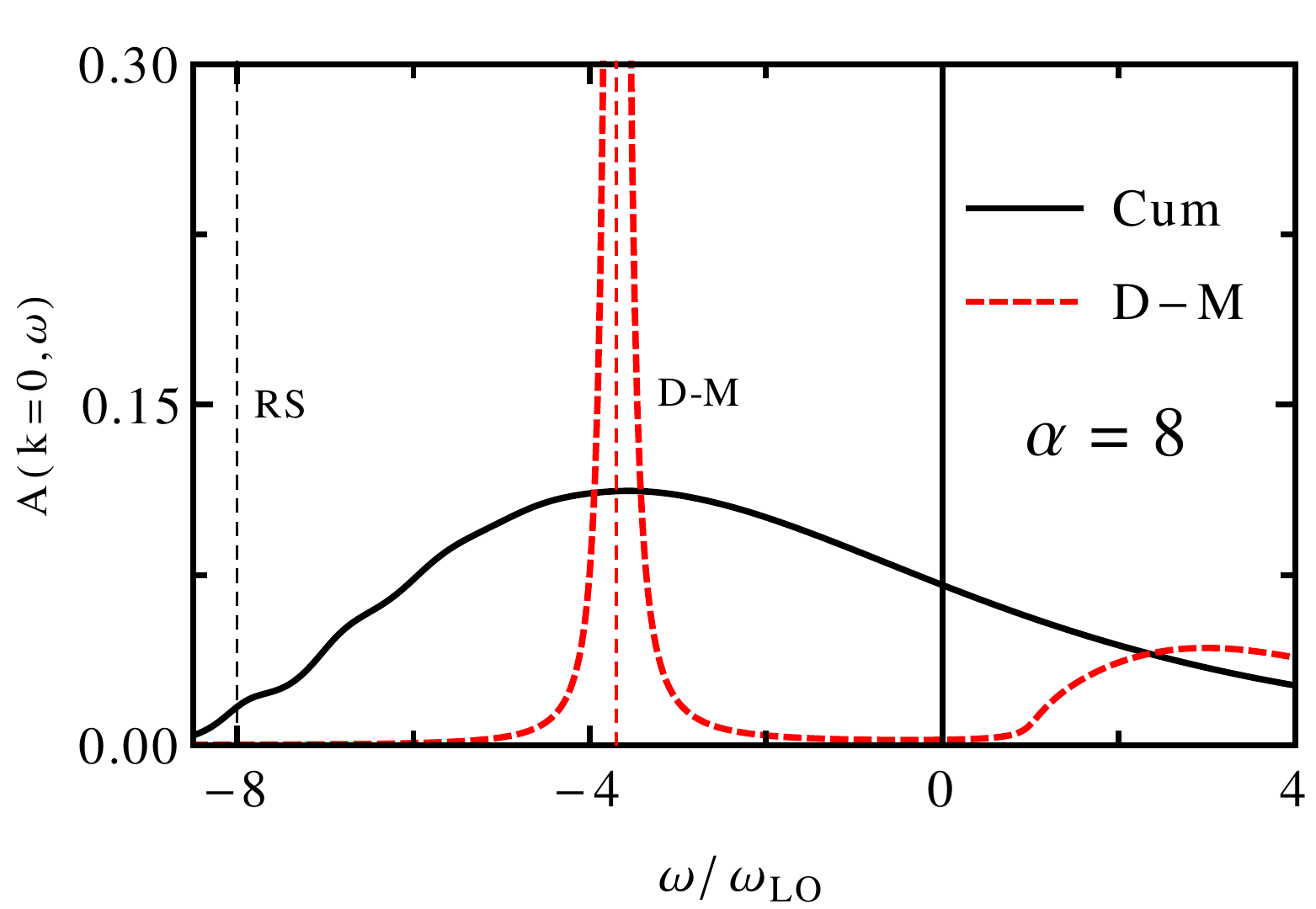}
\caption{Fr\"ohlich Hamiltonian spectral function using the cumulant $G_C$ 
(in black) and the Dyson-Migdal approach $G_D$ (dashed, in red) for $\alpha$=8, from the 
Migdal self energy broadened by $\delta=0.12\omega_{LO}$. \label{spectral_functions_alpha8}}
\end{figure}

Let us analyze the behavior of the quasiparticle peak in the D-M case in more detail, in the case
without any numerical broadening.
The quasiparticle energy is found from
\begin{equation}
E^{D}_{QP}=\Re e \, \Sigma(E^{D}_{QP}).
\label{eq:QP}
\end{equation}
Using Eq.\eqref{eq:ReSigmaFrohlich_e} for electron states gives a cubic equation,
\begin{equation}
\alpha^2=\left(\frac{E_{QP}^D}{\omega_{LO}}\right)^2-\left( \frac{E_{QP}^D}{\omega_{LO}}\right)^3; \ \ (E_{QP}^D < 0).
\label{eq:cubic}
\end{equation}
For all $\alpha$, this has one negative real root; $E_{QP}^D\sim -\alpha\omega_{LO}$ for small $\alpha$
and $E_{QP}^D\sim -\alpha^{2/3} \omega_{LO}$ for large $\alpha$.  This agrees with Fig. \ref{fig:QPenergy} (red circles).
For small $\alpha$, the leading correction to RS is
\begin{equation}
E^{D}_{QP}\approx-(\alpha-\alpha^2/2)\omega_{LO}
\label{eq:QP1}
\end{equation}
The D-M quasiparticle energy Eq.(\ref{eq:QP}) corresponds to the intersection of the $\Re e \, \Sigma(\omega)$ function
with the straight $\omega$ line, as shown in Fig.~\ref{spectral_functions_alpha1}.
As shown in Fig.~\ref{fig:QPenergy}, the second-order Rayleigh-Schr\"odinger answer 
$E_{QP}^{RS}=-\alpha\omega_{LO}$, although not perfect, is better than the 
 Dyson-Migdal answer (with $\Sigma_M=G_0 D$, not the self-consistent $\Sigma_M=GD$, see Ref.\onlinecite{note-selfconsistency}).

The Fr\"ohlich spectral function $A=-\Im m \, G_R(\omega)/\pi$, 
in D-M approximation, 
has two parts.  The quasiparticle part is $Z^{D}\delta(\omega-E^{D}_{QP})$,
where $Z^{D}=1/(1-d\Sigma/d\omega)$ is evaluated at $\omega=E^{D}_{QP}$.  To lowest order, the quasiparticle
weight $Z^{D}\approx 1-\alpha/2$. At large $\alpha$,
$Z^{D}$ tends asymptotically to $2/3$, but the linearized weight, Eq.(\ref{eq:ZDMlin})
is $Z^{Dlin}\approx 1/(1+\alpha/2)$ for all values of $\alpha$,
which tends to zero asymptotically.
The D-M spectral function (without numerical broadening) is
\begin{equation}
A(x)=\frac{Z}{\omega_{LO}} \delta(x-x_{QP})
+\frac{1}{\pi\omega_{LO}}\frac{\alpha\sqrt{x-1}}{x^2 (x-1)+\alpha^2}\theta(x-1),
\label{eq:A1}
\end{equation}
where $x=\omega/\omega_{LO}$ and $x_{QP}=E^{D}_{QP}/\omega_{LO}$.
The D-M side-band  
always starts at $\omega=\omega_{LO}$, rather than
at the intuitively correct value of $E_{QP}+\omega_{LO}$.  The Monte Carlo spectral functions
\cite{Mishchenko2000} show sidebands starting close to the intuitive energy.

%---------------------------------------------------------

\section{FULL MIGDAL SELF-ENERGY FROM FIRST-PRINCIPLES}

We present now first-principles results for the full self energies 
(real and imaginary parts, from all phonon modes, including
interband and Debye-Waller effects) of MgO and LiF, at the valence band maximum (VBM)
and conduction band minimum (CBM).  
These results will be used in the next section to find D-M and cumulant spectral functions.
This section also gives
the related first-principles parameters to be fed into the Fr\"ohlich model.
For MgO and LiF band extrema, the  Fr\"ohlich coupling $\alpha$ ranges from about 1.5 to 15.
%as we will see.
We also tabulate the magnitudes of the separate Debye-Waller and Fan terms, 
as well as their contributions from unoccupied and occupied states.

Technical details of the first-principles calculations are in the Appendix.
The most delicate issue concerns the sampling of phonon wavevectors in the Brillouin zone, and 
the numerical broadening factor needed to treat Eq.(\ref{self_energy_Fan}). 
To obtain well-converged self energies, the eigenenergy differences between sampled wavevectors
as well as the numerical broadening factor must be significantly smaller than the LO phonon frequency.
This is especially important at the unperturbed quasiparticle eigenenergy, where the real and imaginary 
parts and their derivatives govern the asymptotics of the cumulant, and hence the quasiparticle peak characteristics.
We choose a broadening of 0.01 eV, approximately $\omega_{LO}/8$ (see Table \ref{tab:basicprop}), 
and wavevector grids up to 96 $\times$ 96 $\times$ 96 points for the CBM of MgO and 
48 $\times$ 48 $\times$ 48 points for the other cases.
This is considerably better than in Ref.~\onlinecite{Antonius2015} for the same materials
(diamond and BN were also studied in that work). 
Indeed, in Ref.~\onlinecite{Antonius2015}, the broadening factors ranged between 0.1eV and 0.4eV, 
and phonon wavevector grids had at most 32 $\times$ 32 $\times$ 32 points.

MgO and LiF both crystallize in the (cubic) rocksalt structure, with one formula unit per primitive cell. 
Density-Functional Theory (DFT-GGA) Kohn-Sham electronic structure of both materials
can be found elsewhere\cite{Antonius2015}, and will not be reproduced here.
In both materials, the CBM is not degenerate. It is parabolic in a large region
around $\Gamma$, so we expect the effective mass parabolic approximation 
to be adequate. The VBM is triply degenerate at $\Gamma$. One light hole band 
rapidly separates from two heavy hole bands away from $\Gamma$, 
the latter being degenerate along the $\Gamma$-X and $\Gamma$-L directions. 
The deviation with respect to parabolicity is faster than for the conduction band. 
The VBM eigenenergy is set to zero by convention. 
Concerning the phonon band structures, in both materials,
there are three acoustic and three optic phonon branches.
At $\Gamma$, the LO branch is separated from the doubly-degenerate TO branches. 

Table \ref{tab:basicprop} presents the computed 
geometric, electronic, dielectric, and dynamical properties of MgO and LiF,
that determine the corresponding Fr\"ohlich parameter $\alpha$, also reported in this table.
The primitive cell parameter is $a_0=a/\sqrt{2}$, where $a$ is the size of the conventional cube.
Different effective masses
are mentioned for the valence bands, corresponding 
to the heavy hole (hh) and to the light holes (lh), and also 
to different directions of the non-spherical electronic structure \cite{Mecholsky2014}.
Note that the dynamical properties of the two materials 
are rather similar, while their dielectric properties and effective masses differ significantly.
The Fr\"ohlich parameter $\alpha$ %can be deduced, and can 
provides a rough estimate
of the phonon-induced zero-point renormalization of the quasiparticle energy.
For the conduction band minimum, the estimated shifts (-$\alpha_e\omega_{LO}$) are -0.137eV for MgO
and -0.332eV for LiF. 
For valence bands, we do not attempt to integrate over all 
effective mass directions and hole types, but simply provide the corresponding 
$\alpha$ values deduced from Eq.(\ref{eq:alpha}).
The values presented in Table I are in reasonable agreement with those recently computed
in Ref.~\onlinecite{Lambrecht2017} for the same materials. 
However, in the latter work, the
Fr\"ohlich polaron binding energy is defined as half the value from the usual theoretical approach 
(that we adopt),  because the authors cut off the q-integral at $\pi$ over the polaron radius instead of infinity. 
We find on the contrary that the
Fr\"ohlich values underestimate first-principles values, as will appear later.

\begin{table}[h]
\caption{\label{tab:basicprop}Computed basic characteristics of MgO and LiF. See text for the different symbols.
}
\begin{tabular}{l c | c | c}
 \hline \hline
& Unit & MgO & LiF \\ 
$a_0$ & [\AA] & 3.01 & 2.88 \\
$\Omega_0$ & [\AA$^3$] & 19.2 & 16.9 \\
$\varepsilon_g$ (DFT-GGA) & [eV] & 4.49 & 8.54 \\
$\omega_{LO}$/$\omega_{TO}$ & [eV] & 0.0844/0.0454 & 0.0828/0.0466 \\
$\epsilon_\infty$/$\epsilon_0$ & & 3.23/11.14 & 2.04/6.44\\
m$^\ast_e$ & & 0.340 & 0.873 \\
m$^\ast_{hh}$ ($\Gamma$-X/$\Gamma$-L) & & 2.164/3.822 &  3.622/11.955 \\
m$^\ast_{lh}$ ($\Gamma$-X/$\Gamma$-L) & & 0.387/0.335 & 1.346/0.887  \\
$\alpha_e$ &  & 1.624 & 4.009 \\
$\alpha_{hh}$ ($\Gamma$-X/$\Gamma$-L) & & 4.101/5.450 & 8.165/14.834 \\
$\alpha_{lh}$ ($\Gamma$-X/$\Gamma$-L) & & 1.734/1.610 & 4.977/4.040 \\
\hline \hline
\end{tabular}
\end{table}

\begin{figure}
\centering
\includegraphics[width=0.45\textwidth]{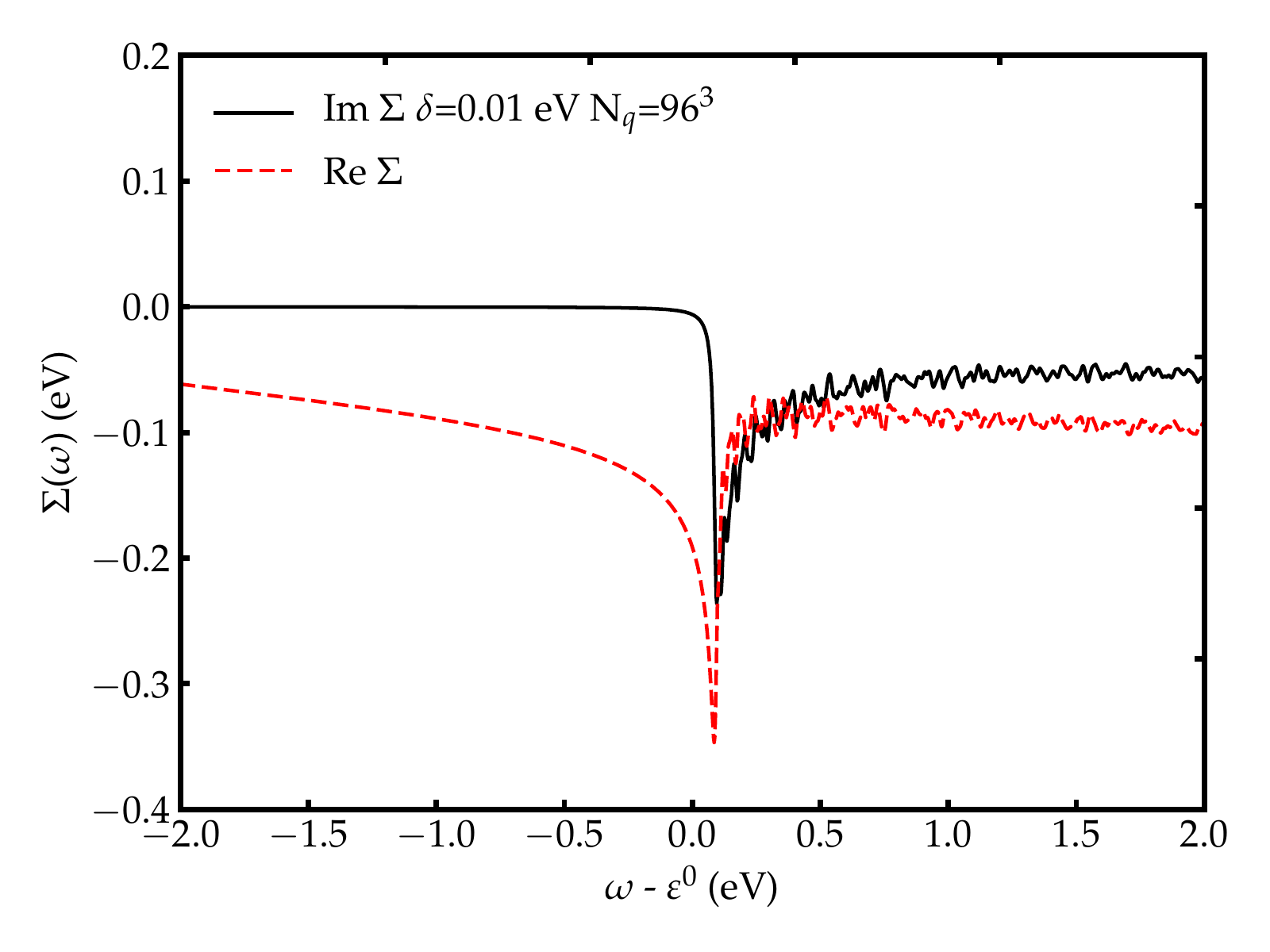}
\caption{The MgO conduction band minimum retarded self energy with $\delta$ = 0.01 eV and a 96$^3$ q-grid. Full black line: imaginary part;
dotted red line: real part.  The reference energy $\varepsilon^0$ is the unrenormalized conduction band energy minimum. This figure agrees with Fig. 6, except that it includes all phonons and interband effects rather than just the analytic Fr\"ohlich result.
\label{fig:MgO_CBM_slf_001_96x}
}
\end{figure}

\begin{figure}%[top]
\includegraphics[angle=0,width=0.45\textwidth]{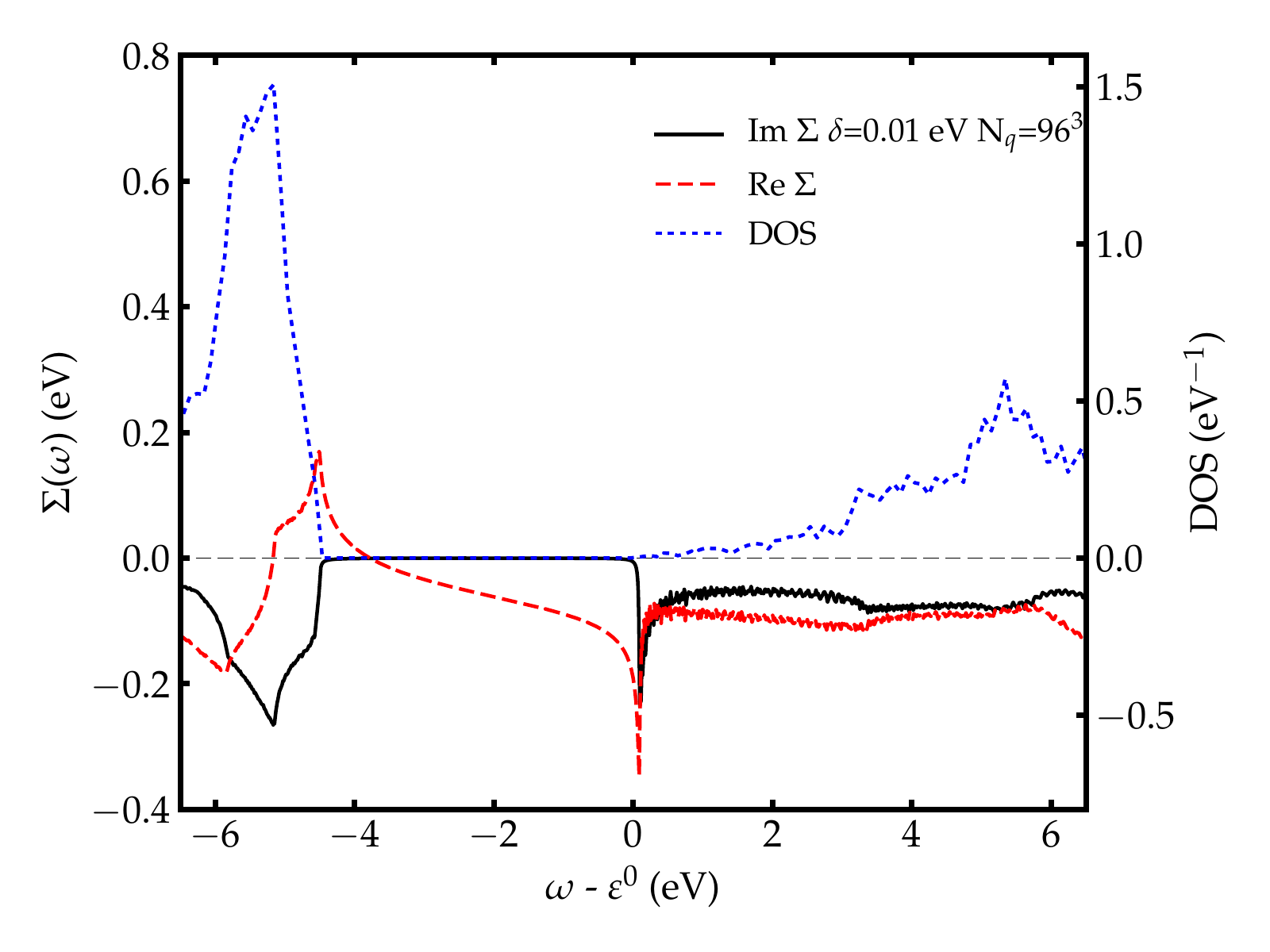}
\caption{Retarded self energy for the bottom of the conduction band of MgO in a wider range of energy than in Fig. \ref{fig:MgO_CBM_slf_001_96x}: imaginary part in black, real part in dashed red.  The electronic DOS is also shown  (dotted blue), for comparison. }
\label{fig:MgO_CBM_slf}
\end{figure}

%We now focus on MgO, and analyze its self energy for both the CBM and VBM.
The self energy for the CBM of MgO, in a 2eV window
around the bare electronic energy $\epsilon^0$, is presented in Fig. \ref{fig:MgO_CBM_slf_001_96x}.
For sake of brevity, we introduce the notations $\Sigma_1= \Re e \Sigma$ and $\Sigma_2= \Im m \Sigma$.
%In the vicinity of the DFT CBM energy, the structure of the self energy is 
%very similar to the Fr\"ohlich one, see Fig.~\ref{spectral_functions_alpha1}. 
Fr\"ohlich-type real and imaginary peaks, both negative, occur at $\omega_{LO}=0.0844 eV$,
just as in Fig. \ref{spectral_functions_alpha1}.  
Despite a very fine 96$^3$ q-point grid sampling and a small 0.01 eV broadening of the denominators (see the Appendix),
some numerical noise is still clearly visible.
The Debye-Waller contribution, and the Fan contributions from bands other than lowest conduction,
shift $\Sigma_1$ to more negative values compared to Fig. \ref{spectral_functions_alpha1},  
and give it a slight linear slope in the vicinity of the conduction band minimum.
This is shown in Fig. \ref{fig:MgO_CBM_slf}, that presents the same data in a wider energy window,
also with the electronic density of states (DOS).
Other structures are indeed present, in the valence band region (below -4 eV), with the 
same van Hove singularities as the electronic DOS. Small structures in the conduction band region are seen as well.

\begin{figure}%[top]
\includegraphics[angle=0,width=0.45\textwidth]{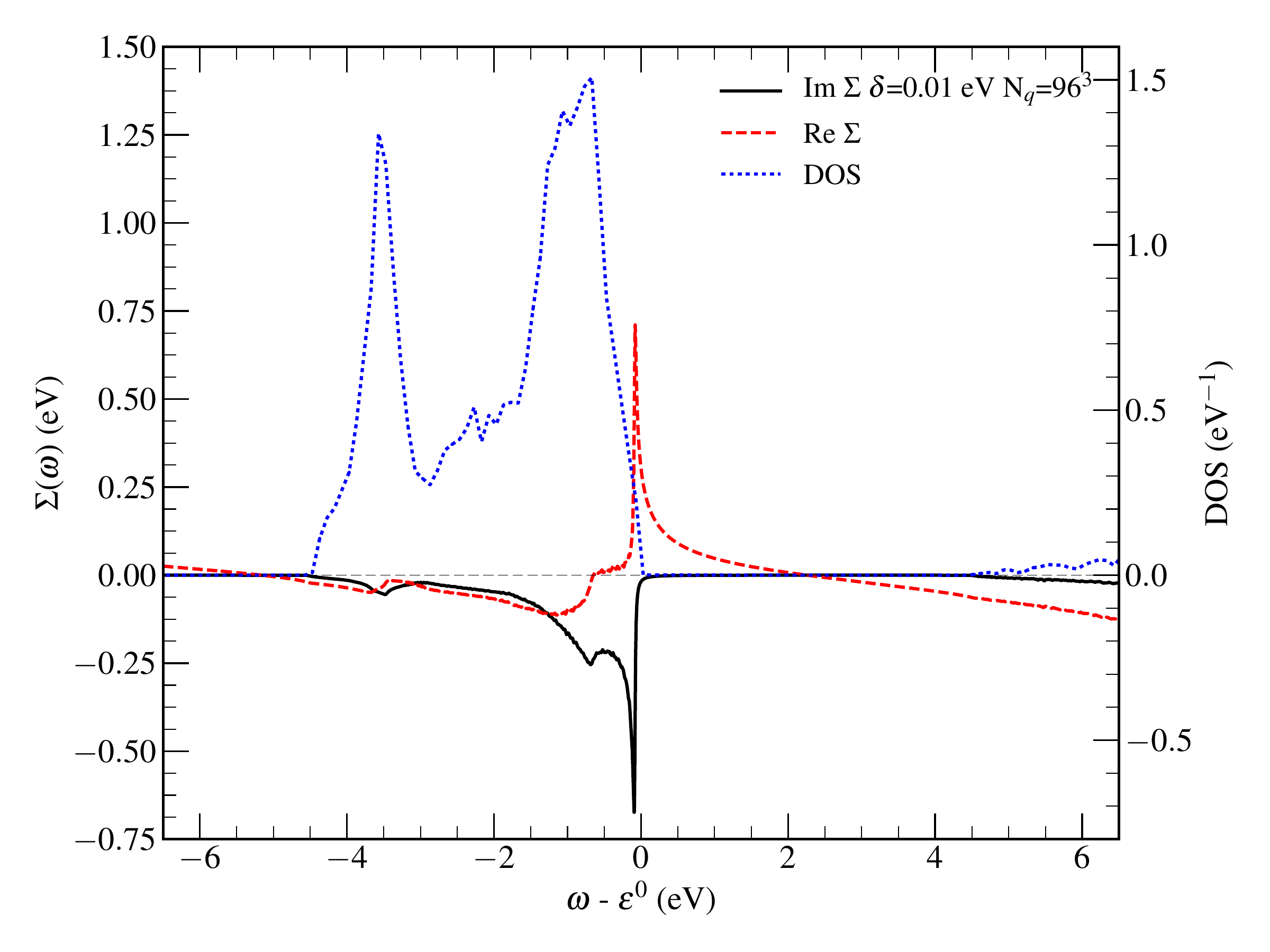}
\caption{Retarded self energy for the top of the valence band of MgO: imaginary part in black, real part in dashed red. 
The electronic DOS is also shown (dotted blue), for comparison. The valence band minimum eigenenergy is 
the reference energy $\varepsilon^0$, which explains the horizontal shift of the DOS with respect to 
Fig. \ref{fig:MgO_CBM_slf}.}
\label{fig:MgO_VBM_slf}
\end{figure}

Similarly, $\Sigma_1$ and $\Sigma_2$ for the VBM of MgO are shown in
Fig. \ref{fig:MgO_VBM_slf}. With the same 
sampling and broadening as for the CBM, the noise can hardly be seen.
Indeed, the curvature of the hole band is less pronounced for the VBM than for the CBM,
making the numerical work less difficult.
Close to $\omega-\epsilon^0=0$, the structure of the self energy is close to the Fr\"ohlich self energy,
with the appropriate sign change for a hole polaron. 
Additional valence band characteristic features are seen, clearly related to the electronic DOS.
By contrast, the imaginary part of the self energy in the conduction region is very 
small, and the real part is nearly structureless.

Unlike the electronic dispersion, 
the role of the phonon dispersion is apparently minor. Indeed, 
for the phonon frequencies to have an impact on the self energy,
the difference between $\omega \approx \epsilon_{\mathbf{k}n}$ and the electronic eigenenergies $\epsilon_{\mathbf{k+q}n'}$ 
must be comparable to phonon frequencies,
see Eqs.(\ref{self_energy_Fan_un}) and (\ref{self_energy_Fan_oc}). 
This happens only in a small Brillouin zone volume around $\Gamma$, 
in which the phonon branches are  practically constant.


\begin{table}[h]
\caption{\label{tab:selfenergy}MgO and LiF first-principles self energy (eV) and frequency derivative 
(dimensionless) at $\omega=\varepsilon_{\mathbf{k}n}$, and their components,
for the conduction band minimum and valence band maximum. 
The Debye-Waller self energy is static (frequency-independent) and real.
The quasiparticle weights, from linearized D-M and retarded cumulant approaches are also mentioned, 
as well as their occupied and unoccupied bands factors. 
For the real part of the self energy, results are reported with two different phonon wavevector grids (96$^3$ and 48$^3$), 
while for the imaginary part, the derivatives and the 
quasiparticle weights, only the results obtained with the 48$^3$ grid are reported. \\
}
\begin{tabular}{l | c | c | c | c}
 \hline \hline
& MgO & MgO & LiF & LiF \\
& CBM & VBM & CBM & VBM \\
 \hline \hline
96$^3$ grid & & & & \\
\hline
$\Sigma_1$ & -0.191 & 0.285 & -0.370 & 0.723 \\
$\Sigma_1^{\mathrm{DW}}$ & -0.056 & 4.263 & 0.078 & 6.785 \\
$\Sigma_{1,un}^{\mathrm{Fan}}$ & -0.371 & -4.327 & -0.524 & -6.911 \\
$\Sigma_{1,oc}^{\mathrm{Fan}}$ & 0.235 & 0.349 & 0.077 & 0.850 \\
 \hline \hline
48$^3$ grid & & & & \\
 \hline
$\Sigma_1$ & -0.175 & 0.285 & -0.342  & 0.695 \\
$\Sigma_1^{\mathrm{DW}}$ & -0.054 & 4.263 &  0.078  & 6.772 \\
$\Sigma_{1,un}^{\mathrm{Fan}}$ & -0.354 & -4.327 & -0.497  & -6.898 \\
$\Sigma_{1,oc}^{\mathrm{Fan}}$ &  0.233 & 0.349 &  0.077  & 0.821 \\
\hline
$\Sigma_2$ & -0.005 & 0.016 & -0.014  & 0.053 \\
$\Sigma_{2,un}^{\mathrm{Fan}}$ & -0.005 & 0.000 & -0.014  & 0.000 \\
$\Sigma_{2,oc}^{\mathrm{Fan}}$ & -0.000 & 0.016 &  0.000  & 0.053 \\
\hline
$\partial\Sigma_1/\partial\omega$ & -0.455 & -1.594 & -1.353  & -4.780 \\
$\partial\Sigma_{1,un}^{\mathrm{Fan}}/\partial\omega$ & -0.446 & -0.007 & -1.345  & -0.012 \\
$\partial\Sigma_{1,oc}^{\mathrm{Fan}}/\partial\omega$ & -0.009 & -1.587 & -0.008  & -4.768 \\
\hline
$\partial\Sigma_2/\partial\omega$ & -0.058 & -0.252 & -0.206  & -1.248 \\
$\partial\Sigma_{2,un}^{\mathrm{Fan}}/\partial\omega$ & -0.058 & 0.000 & -0.206  & 0.000 \\
$\partial\Sigma_{2,oc}^{\mathrm{Fan}}/\partial\omega$ &  0.000 & -0.252 & 0.000  & -1.248 \\
\hline
$Z^{Dlin}_{\mathbf{k}n}$ & 0.687 & 0.386 & 0.425  & 0.173 \\
$Z^R_{\mathbf{k}n}$ &  0.634 & 0.204 & 0.258  & 0.008 \\
$Z^{un}_{\mathbf{k}n}$ &  0.640 & 0.993 &  0.260  & 0.988 \\
$Z^{oc}_{\mathbf{k}n}$ &  0.991 & 0.205 &  0.992  & 0.008 \\
\hline \hline
\end{tabular}
\end{table}
%

%
%\begin{table}[h]
%\caption{\label{tab:selfenergy_OLD} OLD table. MgO and LiF first-principles self energy (eV) and frequency derivative %(adimensional) at $\omega=\varepsilon_{\mathbf{k}n}$, and their components,
%for the conduction band minimum and valence band maximum. The Debye-Waller self energy is static (frequency-independent) %and real.
%The quasi-particle weights, from D-M and retarded cumulant approaches are mentioned, as well as their occupied and %unoccupied bands factors. \\
%}
%\begin{tabular}{l | c | c | c | c}
% \hline \hline
%& MgO & MgO & LiF & LiF \\ 
%& CBM & VBM & CBM & VBM \\ 
% \hline \hline
%$\Sigma$ & -0.160-i0.003 & 0.269+i0.015 & -0.316-i0.011  & 0.667+i0.050\\
%$\Sigma^{DW}$ & -0.059 & 4.316 & 0.078 & 6.771 \\
%$\Sigma^{FAN}_{un}$ & -0.342-i0.003 & -4.381+i0.000 & -0.471-i0.011 & -6.987+i0.000\\
%$\Sigma^{FAN}_{oc}$ & 0.241+i0.000 & 0.334+i0.015 & 0.076+i0.000 & 0.793+i0.050\\
%\hline 
%$\frac{\partial\Sigma}{\partial\omega}$ & -0.309-i0.029 & -1.452-i0.222 & -1.071-i0.143 & -4.417-i1.175\\
%$\frac{\partial\Sigma^{FAN}_{un}}{\partial\omega}$ & -0.300-i0.029 & -0.007+i0.000 & -1.063-i0.143 & -0.012+i0.000\\
%$\frac{\partial\Sigma^{FAN}_{oc}}{\partial\omega}$ & -0.009+i0.000 & -1.444-i0.222 & -0.008+i0.000 & -4.405-i1.175\\
%\hline 
%$Z^D_{\mathbf{k}n}$ & 0.764 & 0.408 & 0.483 & 0.185 \\
%$Z^R_{\mathbf{k}n}$ & 0.735 & 0.234 & 0.343 & 0.012 \\
%$Z^{un}_{\mathbf{k}n}$ & 0.741 & 0.993 & 0.445 & 0.988 \\
%$Z^{oc}_{\mathbf{k}n}$ & 0.991 & 0.236 & 0.992 & 0.012 \\
%\hline \hline
%\end{tabular}
%\end{table}
%

The characteristics of the self energy, evaluated at the bare eigenenergy $\varepsilon_{\mathbf{k}n}$, 
are reported in Table \ref{tab:selfenergy},
including the decomposition into Debye-Waller and Fan (and unoccupied/occupied) components. 
Also, the frequency derivative of the self energy and components are given.
A $48 \times 48 \times 48$ q-wavevector grid has been used by default for this table,
except for the real part of the self energies and their decomposition, which is also given
using the more converged $96 \times 96 \times 96$ q-wavevector grid.

The convergence of the Allen-Heine-Cardona zero-point renormalization (ZPR)
with respect to the wavevector sampling has been thoroughly analyzed in Sec. IV.B.2 of 
Ref.~\onlinecite{Ponce2015}. In particular,
for IR active materials treated in the non-adiabatic approximation, at the band structure extrema, a $1/N_q$ behavior
is obtained, where $N_q$ is the linear density of q-points of the three-dimensional sampling. As shown in the appendix, 
Table III, such a trend matches well numerical results. Thus, the $48^3$ and $96^3$ grids $\Sigma_1$
results can be extrapolated to infinity, giving for the CBM and the VBM MgO, respectively, a ZPR
(or polaron binding energy) of $-207$meV and $319$meV,
and for the CBM and VBM of LiF, respectively, a ZPR of $-398$meV and $751$meV.
The total band gap ZPR for MgO is $526$meV while for LiF it is $1149$meV.

The Fr\"ohlich estimated CBM shifts (-$\alpha_e\omega_{LO}$), {\it i.e.} $-137$meV for MgO
and $-332$meV for LiF, are in qualitative agreement with first-principles results, 
but underestimate their absolute value by about $50$-$70$meV.
Still, the first-principles ZPR of the CBM in these materials is 
thus apparently largely dominated by the Fr\"ohlich part of the electron-phonon interaction. The analysis of the VBM shift is more complex due to the band warping, and will not be  given here. Still, the range of $\alpha$ for holes mentioned in Table I and the zero-point renormalisation for the VBM in Table II are quite consistent. 
A similar dominance of the 
Fr\"ohlich part of the electron-phonon interaction in other infra-red active materials with large LO-TO splitting is expected,
and would be consistent with the widespread use of the Fr\"ohlich Hamiltonian for the interpretation of many experimental results.

Our self energy values compare favorably with those of 
Table I, column $\Sigma^{dyn}(\varepsilon^0)$ of Antonius et al. \cite{Antonius2015}.
Remember however, that in the latter study, the broadening factor $\delta$ was much larger 
and the sampling of phonon wavevectors
much coarser than in the present study, see the Appendix.  Actually, the quantities reported in the 
column $\Sigma^{stat}(\varepsilon^0)$ of Table I of Ref.~\onlinecite{Antonius2015} should diverge for vanishing
broadening factor and perfect Brillouin Zone sampling, for the IR-active materials 
BN, MgO and LiF, as shown in Ref.~\onlinecite{Ponce2015}. The similarity of 
$\Sigma^{stat}(\varepsilon^0)$ and $\Sigma^{dyn}(\varepsilon^0)$ is thus an artifact, 
simply due to the similarity of the chosen broadening factor value ($\delta=0.1$ eV) with the LO phonon frequency
in these materials (see {\it e.g.} Table \ref{tab:basicprop}). 

The decomposition of $\Sigma_1$ into its Debye-Waller and Fan components 
highlights the dramatic cancellation
between the Debye-Waller component and the unoccupied bands Fan components, for the VBM of the two materials.
As a consequence, the occupied band Fan component has the same magnitude as the total zero-point renormalization value.  
By contrast, the CBM zero-point renormalization comes from contributions with different signs, without noticeable cancellation. The emergence of
a total shift given quite accurately by the Fr\"ohlich approach is rather surprising, in view of such data. The sum rule
for acoustic modes, presented in Ref.~\onlinecite{Allen1976}, is without doubt at play 
in the final dominance of the Fr\"ohlich estimation.

In Table \ref{tab:selfenergy}, we also report the quasi-particle weights, from linearized D-M and retarded cumulant approaches, 
that are directly obtained from the derivative of the real part of the self energy with respect to 
the frequency at the bare electronic energy, see  Eqs. (\ref{eq:ZDMlin}) and (\ref{eq:ZR}).
The {\it retarded} cumulant weights can be decomposed in their hole and electron factors, 
following Eqs.(\ref{eq:self_energy_Fan_e+h}) and (\ref{eq:ZR}):
\begin{equation}
Z^R_{\mathbf{k}n}= Z^{un}_{\mathbf{k}n}Z^{oc}_{\mathbf{k}n}.
\label{eq:Zeh}
\end{equation}
%
%\stkout{Only one of these factors is included in the t-O cumulant weights of Refs.~
%\onlinecite{Guzzo2011,ZhouThesis2015,Gumhalter2016}, 
%while both are present in Ref.~\onlinecite{Aryasetiawan1996}.
%In the latter t-O approach.}
%%%%
Only one of these factors is included in the t-O cumulant
weights of Refs. \onlinecite{Aryasetiawan1996}, \onlinecite{Guzzo2011}, \onlinecite{Gumhalter2016},
and \onlinecite{ZhouThesis2015} (Eq. (3.64)), while both are present in
Ref. \onlinecite{ZhouThesis2015} (Eq. (2.62)).  In the former t-O approaches,
the imaginary part of the self energy (which comes from the electron or hole-only self energy)
is not consistent with the asymptotic limit of the cumulant, determined from the complete self energy expression,
so that the spectral function is not normalized. However, as can be judged by the closeness to unity 
of $Z^e_{\mathbf{k}n}$ in the VBM case and $Z^h_{\mathbf{k}n}$ in the CBM case, in MgO and LiF, the 
normalization defect is very small: the smallest of these weights is at least 0.988, that is only 1.2\% less than one. 
%%%%
%\stkout{Thus, for the present study, the different cumulant approaches are nearly equivalent.  
%This lack of impact of the unoccupied states on the VBM self energy, and of the occupied states on the 
%CBM self energy can be traced back to the large ratio between the 
electronic gap and the largest phonon frequency.
%This might not be true for small-gap semiconductors.}
%%%%
% This discussion is now inserted after Eqs. 30-34.}

Unlike $Z^R_{\mathbf{k}n}$,  $Z^{Dlin}_{\mathbf{k}n}$ cannot be factorized in a product of occupied and unoccupied state contributions, see Eq.~\ref{eq:ZDM}.
$Z^{Dlin}_{\mathbf{k}n}$ and $Z^R_{\mathbf{k}n}$ spectral weights differ the most in the VBM case.
For the LiF VBM case, the ratio exceeds one order of magnitude. It is slightly less than two for the MgO VBM.
Taking into account the results from large polaron studies of the Fr\"ohlich Hamiltonian (clearly favoring the
Rayleigh or cumulant shifts, Eq.(\ref{eq:ERS}) and (\ref{eq:ER})), 
the values from Table I of Ref.~\onlinecite{Antonius2015}, column $\Sigma^{dyn}(\varepsilon^0)$ 
have also to be preferred over the
values in columns $Z\Sigma^{dyn}(\varepsilon^0)$ or $\Sigma^{dyn}(\varepsilon)$. 
%By the way, 
Similarly the values published
in Ref.~\onlinecite{Ponce2015}, Table VII, column ``ZPR Non-adiabatic" correspond to the preferred expressions Eq.(\ref{eq:ERS}) and (\ref{eq:ER}). 

For completeness, Table \ref{tab:selfenergy} also mentions the imaginary values of the self energy and its 
derivative, respectively, linked physically to the broadening of the quasiparticle peak and its asymmetry. 
Note that these values are actually artificial effects of numerical broadening. 
Our computations only include band extrema at zero temperature, for which the imaginary 
part of Eqs.~\ref{self_energy_Fan_un}
and~\ref{self_energy_Fan_oc} vanish exactly if there is no artificial broadening. 
Nonzero values of $\Im m\Sigma$ at the QP energy thus indicate the magnitude of the broadening parameter. 

%---------------------------------------------------------
\section{FIRST-PRINCIPLES SPECTRAL FUNCTIONS}
 
%In this section, we compute 
Here we present the spectral functions of the CBM and VBM of MgO and LiF 
using $G_C$ and $G_D$, and examine their differences.
The  LiF VBM spectral function $A_{\rm DM}$ from $G_D$ was previously given in 
Fig. 1 of Ref.~\onlinecite{Antonius2015}, but our numerical treatment of $G_D$ is significantly improved.
%with respect to this work. 

We only use the retarded cumulant approach. It appears to be the preferred method 
to  obtain the quasiparticle spectral functions for insulators as well as metals. The previous section has described, however, the
similarity between retarded and time-ordered cumulants for our gapped
materials, due to the smallness of the contribution of the occupied self energy to
the CBM, and of the unoccupied self energy to the VBM.

\begin{figure}
\centering
\includegraphics[width=0.45\textwidth]{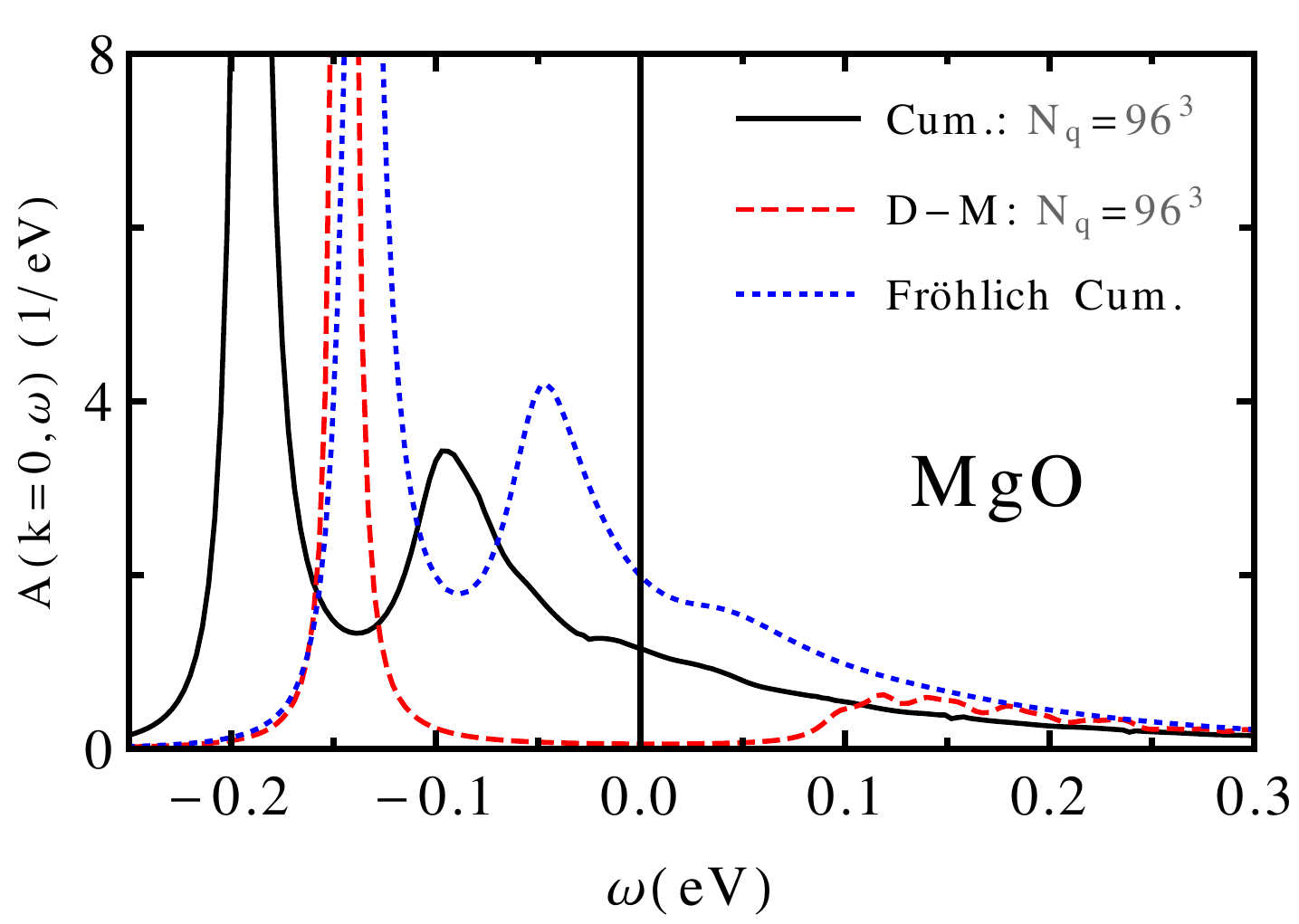}
\caption{Cumulant (black) and Dyson-Migdal (dashed red) spectral functions for the conduction band minimum of MgO. The Fr\"ohlich spectral function (dotted blue) with $\alpha=1.62$ is also shown for comparison.
 \label{fig:MgO_CBM_SF}}
\end{figure}

Figs. \ref{fig:MgO_CBM_SF}
%, \ref{fig:MgO_VBM_SF}, \ref{fig:LiF_CBM_SF} 
to \ref{fig:LiF_VBM_SF} present D-M and retarded cumulant spectral functions, for the CBM and VBM
of MgO and LiF, and also show the Fr\"ohlich spectral function obtained with estimated $\alpha$ (Table II)
for the CBM case.  The phonon wavevector 
Brillouin zone has been sampled by a 96 $\times$ 96 $\times$ 96 grid for the CBM of MgO, and a 48 $\times$ 48 $\times$ 48 grid for the other cases.  In all cases, a 0.01 eV $\approx 0.12 \omega_{LO}$ broadening of the self energy has been used.

For the MgO CBM, with Fr\"ohlich $\alpha=1.62$, Fig.\ref{fig:MgO_CBM_SF}, the position of the quasi-particle 
peak in the first-principles D-M case is lower than from the first-principles cumulant case. Moreover, the 
Fr\"ohlich peak position is closer to the first-principles D-M position than with the 
first-principles cumulant one. This agreement between the D-M position and the
Fr\"ohlich peak position is accidental: the Fr\"ohlich constant is too small to reproduce the
band gap shift from first-principles (cf. the above mentioned $50-70$ meV underestimation), while the D-M shift is also too small, but this is due to the 
incorrect underestimation highlighted in Fig. 1. 
The position of the satellite follows the same pattern as observed 
for the Fr\"ohlich Hamiltonian: the D-M satellite 
is separated from the quasiparticle peak by much more than the LO phonon frequency, 
while the distance between the satellite and the quasiparticle peak in the cumulant case 
is close to the LO phonon frequency value. Hence, we conclude that for the CBM of MgO, 
the spectral function shape is dominated by the LO phonon. First-principles 
and Fr\"ohlich Hamiltonian approaches yield very similar shapes, 
although the Fr\"ohlich approach underestimates the QP energy shift.
This is an important result of the present work.  The same conclusion 
will be obtained for the other band extrema, for both MgO and LiF.

\begin{figure}
\centering
\includegraphics[width=0.45\textwidth]{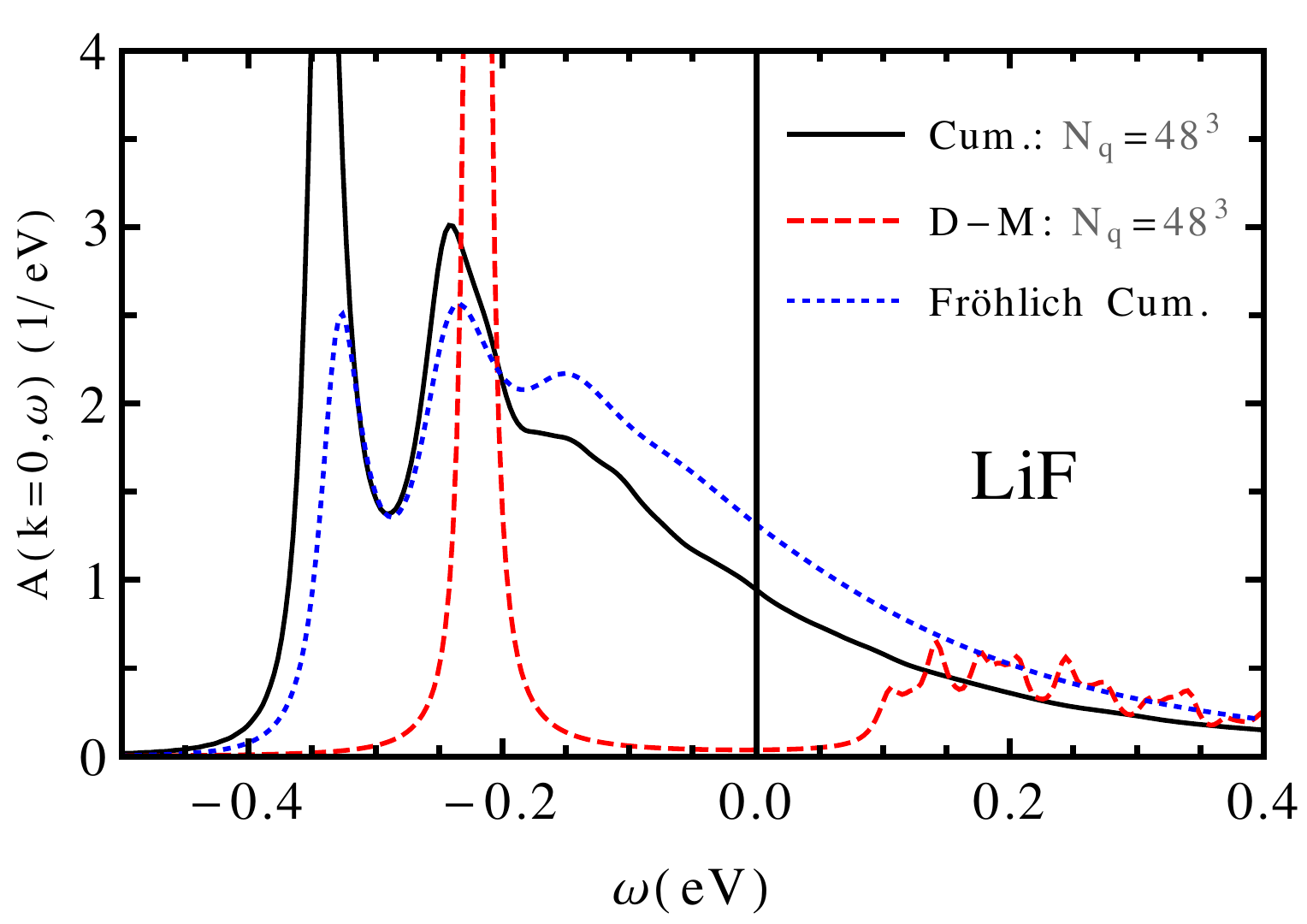}
\caption{Cumulant (black) and Dyson-Migdal (dashed red) spectral functions for the conduction band minimum of LiF.
The Fr\"ohlich spectral function (dotted blue) with $\alpha=4.01$ is also shown for comparison. 
 \label{fig:LiF_CBM_SF}}
\end{figure}

In the LiF CBM case, with Fr\"ohlich $\alpha=4.01$ (Fig.\ref{fig:LiF_CBM_SF}),
the position of the quasi-particle peak in the 
D-M case is much higher than in the cumulant case.
The Fr\"ohlich-only cumulant peak position is close to the full-band cumulant one. 
Because the value of $\alpha$ is larger than for MgO, 
LiF has a larger second satellite before the smoothing of the spectral function. The same 
observations as for the MgO CBM, concerning the shape and position of the peaks,
can also be made. 

\begin{figure}
\centering
\includegraphics[width=0.45\textwidth]{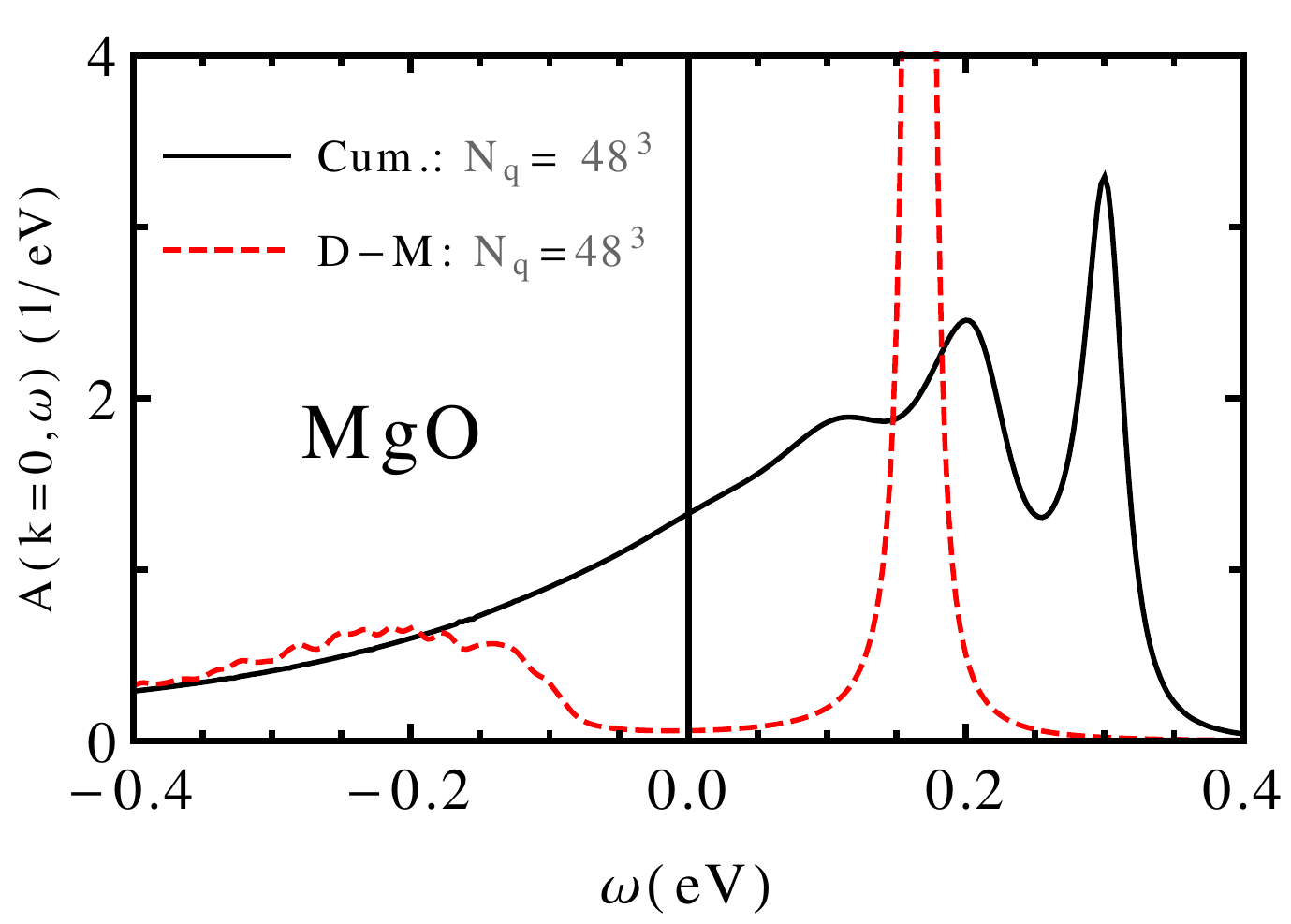}
\caption{Cumulant (black) and Dyson-Migdal (dashed red) spectral functions for the valence band maximum of MgO. 
\label{fig:MgO_VBM_SF}}
\end{figure}

\begin{figure}
\centering
\includegraphics[width=0.45\textwidth]{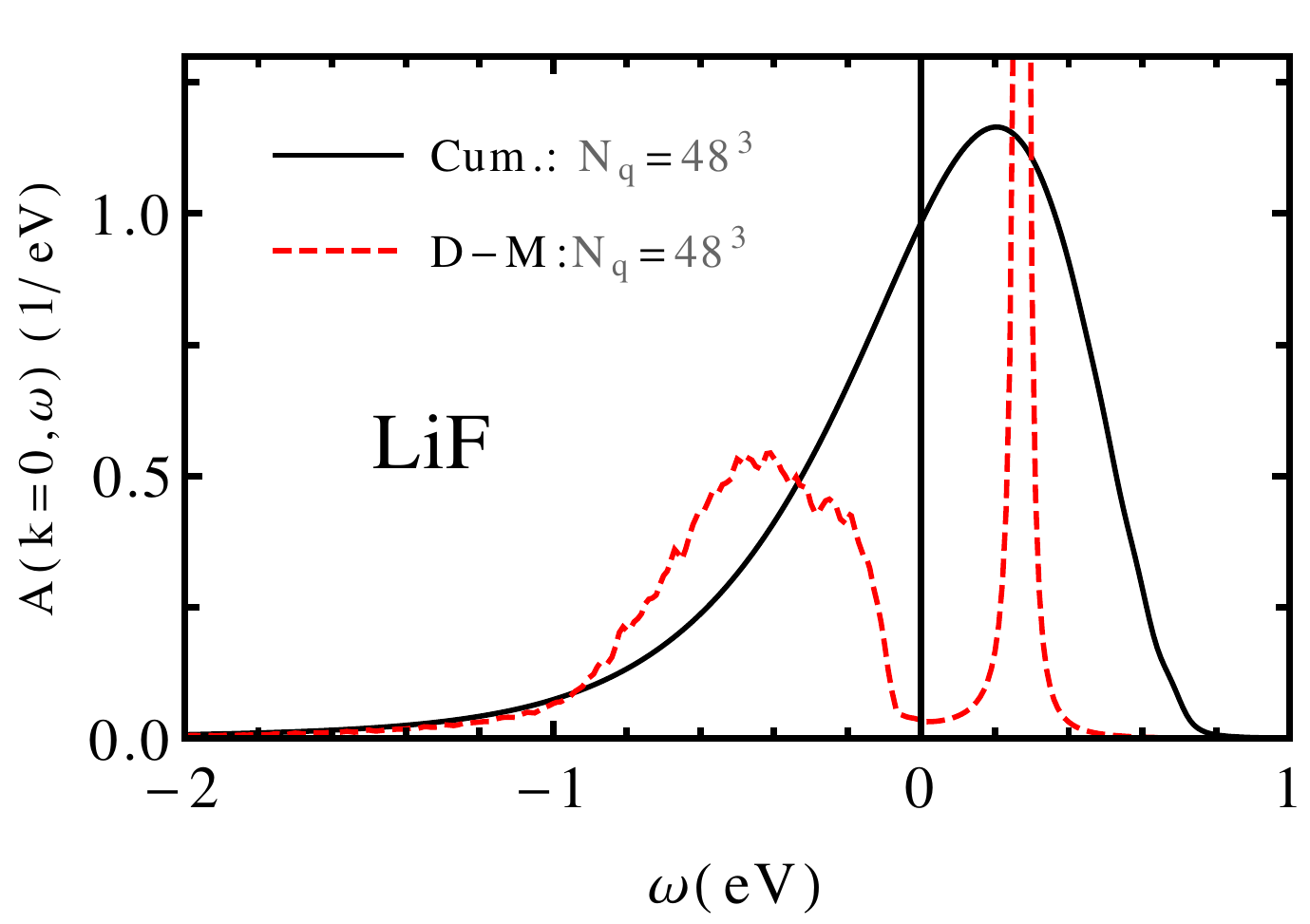}
\caption{Cumulant (black) and Dyson-Migdal (dashed red) spectral functions for the valence band maximum of LiF.
\label{fig:LiF_VBM_SF}}
\end{figure}

Despite the MgO valence band being three-fold degenerate at $\Gamma$, the MgO VBM case, Fig.\ref{fig:MgO_VBM_SF}, is actually very similar to the LiF CBM case, with positive energy shifts seen 
instead of negative energy shifts.

Finally, in the case of LiF VBM, with the Fr\"ohlich $\alpha$ being at most 14.8 for the heavy hole effective mass, Fig.\ref{fig:LiF_VBM_SF}, the cumulant spectral function has become a broad peak 
(similar to Fig. \ref{spectral_functions_alpha8})
without any quasiparticle peak or 
satellite structure, unlike in the D-M case.
The lack of structure in the cumulant spectral function results from the large value of the 
Fr\"ohlich $\alpha$, directly linked to the large hole effective mass, {\it i.e.} the rather flat LiF valence bands
\cite{note-smalldelta}.  


%%%%%%%%%%%%%%%%%%%%%%%%%
\section{SUMMARY AND PERSPECTIVES}
In this work, we compute from first principles the electron-phonon self energy 
for the electronic states at the band extrema of MgO and LiF, at zero temperature, 
from which we obtain the spectral function
using both Dyson-Migdal and the cumulant methodologies. As decades of research 
on model systems emphasize the role of LO phonons in infrared-active semiconductors and insulators,
we first provide an analysis of the self energy and spectral function of the Fr\"ohlich Hamiltonian 
using the same methodologies.

For the imaginary part of the self energy, the characteristic inverse square root behavior
of the Fan self energy found in the Fr\"ohlich Hamiltonian, starting at the LO phonon frequency threshold, 
is also clearly present in MgO and LiF. However, additional structures, mirroring the electronic DOS are also present.
They originate from occupied as well as unoccupied bands, for both electron and hole self energies. 
By contrast, the role of the phonon dispersion is minor.

The real part of the Fr\"ohlich Hamiltonian self energy
has also an inverse square root behavior, %albeit present
on the other side of the LO phonon frequency threshold,
compared to the imaginary part. %Similarly, such 
This feature is also present in first-principles calculations, with additional structures 
mirroring the electronic DOS.  %In this case, however, 
The Debye-Waller self energy, that complements the Fan self energy, gives a global shift, impacting the
zero-point renormalisation of eigenenergies.

To correctly obtain the inverse square root behavior in first-principles self energies, 
the numerical integration over the phonon
wavevectors has to be done carefully, and its convergence monitored.  Numerical broadening of the denominator
present in the Fan self energy helps to reach convergence, but %is detrimental to 
affects the accuracy of the prediction. 
Although schemes to overcome such convergence problems start to appear \cite{Nery2016, Lambrecht2017}, we have
simply performed the summation over a very fine grid, for a small broadening.

The real part of the electron-phonon self energy, evaluated at the bare eigenenergy, is directly linked to the
shift of this bare eigenenergy, in the cumulant approach as well as in the simple Rayleigh-Schr\"odinger approach 
at lowest order. For the Fr\"ohlich Hamiltonian, this correlates well with results from more sophisticated 
methods, for Fr\"ohlich
coupling constants up to about 10.  An estimate of the corresponding Fr\"ohlich coupling constant values
for the conduction band minima gives about 1.62 for MgO and 4.01 for LiF, while for the valence band minima, the situation
is more complex due to band degeneracy and warping, with values in the range 1.6 to 5.5 for MgO, 
and between 4.0 and 14.8 for LiF.

%In this approach, the 
Our first-principles eigenenergy shifts range from $207$ meV for the CBM of MgO to $751$ meV for the VBM of LiF.
%with a reasonable correlation 
Simple Fr\"ohlich Hamiltonian estimates are too low although they account for a large fraction of such shift. 
However, detailed analysis shows that there is cancellation between the Debye-Waller self energy 
(not taken into account in the Fr\"ohlich Hamiltonian) and the Fan self energy. 
This cancellation is especially strong between the unoccupied part of the Fan self energy 
and the Debye-Waller self energy, in the case of the valence band minima for both MgO and LiF.

The derivative of the electron-phonon self energy with respect to the frequency, evaluated at the bare 
eigenenergy, is directly linked to the weight of the quasiparticle peak, in the cumulant approach. 
Different flavors of cumulant methodology are examined, based either on the retarded 
or t-O Green's function. 
For these flavors, the inclusion of occupied and unoccupied band contributions to these weights differ 
in a rather simple way, e.g. the weight is a multiplicative function of the occupied and unoccupied contributions 
in the retarded Green's function approach. %In the present case, 
Due to the large ratio between LO phonon frequency and energy gap, 
the predictions of the different cumulant approaches differ little  
%Indeed, focusing e.g. on 
(the normalisation weights $Z$ differ by less than 2\%).
The difference might be larger for materials with smaller gaps and larger maximum phonon frequencies, 
as it will scale roughly 
%proportionaly to 
as the ratio of these quantities.

We analyze detailed structure of spectral functions from both cumulant and Dyson-Migdal approaches,
%also with representative 
using both the Fr\"ohlich Hamiltonian and full DFT.
%spectral functions from the cumulant approach. 
In the cumulant case, the separation between the quasiparticle peak and the first satellite 
corresponds to the LO phonon frequency. For equivalent Fr\"ohlich coupling constant 
equal to about four, one also observes a second satellite, separated from the first one 
by the 
LO phonon frequency as well. Beyond these, the cumulant spectral function decays monotonically 
and any other structure is hard to discern, although this is impacted by the numerical (small) broadening that we have used. 
This behavior is very similar to that of the cumulant Fr\"ohlich spectral function. 
For the LiF valence band maximum, the equivalent Fr\"ohlich coupling constant 
is higher than four, and the spectral function is a broad peak, showing no quasiparticle peak. 
The behavior of the LiF and MgO spectral functions at valence and conduction 
band extrema in D-M theory differs strongly from that of the cumulant approach. In the former, 
there is always a clear quasiparticle peak, as well as a broad satellite feature, 
whose starting energy differs by more than the LO phonon frequency from the quasiparticle peak. 
This behavior is seen
%mimics the one encountered 
in the Fr\"ohlich Hamiltonian case as well. 
Since the quasiparticle peak location is badly predicted from Dyson-Migdal theory 
even at low Fr\"ohlich coupling constants, 
%there is reason to believe 
it is safe to argue that the shape of spectral function 
from the Dyson-Migdal approach can never be trusted.

%We believe that
Our conclusions should apply to a large class of infrared-active materials. In particular, 
the polar LO phonon contribution will give the biggest part of the zero-point renormalization
and the spectral lineshape. 
It would be worth to examine more materials, and find a rule
for when the
Fr\"ohlich Hamiltonian might give a reasonably accurate estimate of the electron-phonon quasiparticle shift. 
%In our case, 
For the conduction band minimum, Fr\"ohlich alone accounts for $4/5$ of the CBM shift for LiF, 
but only $2/3$ of the CBM shift for MgO. By contrast, the case of diamond is a
remarkable example of a large QP shift without polar phonons.
Its band edge zero-point renormalisation ($-330$ meV for the indirect gap, and $-416$ meV for the direct gap)
\cite{Ponce2014b, Antonius2014} has a magnitude similar to 
those of the infra-red active materials studied here.

Finally, angle-resolved photoemission (ARPES), although difficult in insulators, could provide a nice test of some
of the predicted spectral functions in this paper.  In particular, consider the MgO valence band maximum shown in
Fig. \ref{fig:MgO_VBM_SF}.  We predict a quasiparticle carrying 20\% of the spectral weight, and clear
phonon satellites.  These may be accessible to experiment.  However, by contrast, consider
the valence band maximum of LiF, shown in Fig. \ref{fig:LiF_VBM_SF}.  The prediction is a 
completely blurred quasiparticle ($Z=0.008$), with a broad peak rather than distinct satellites.  However, 
this prediction is outside the trust range of our approach.  The $\alpha$ values for LiF are in the range 4 to 15, rather than 1.6 to 5.5
for MgO.  Experiment \cite{Woodruff1958} shows that a hole at the top of the valence band of LiF self-traps
in a local distortion similar to an F$_2^-$ ion.  This trapped polaron is known as a ${\rm V_K}$ center 
\cite{Kaenzig1956,Castner1957,Stoneham2001}.  The present simple Hamiltonian (e.g. without anharmonicities) 
is insufficient to yield such a bound state.
The perturbative spectral function correctly tells us that there is negligible weight in the quasiparticle peak.  However, for holes
at the top of the valence band in materials like LiF and NaCl, perturbation theory is not enough.

%%%%%%%%%%%%%%%%%%%%%%%%%
\section*{ACKNOWLEDGMENTS}
We acknowledge helpful discussions with Y. Gillet, A. S. Mishchenko, J.S. Zhou, and J. J. Rehr.  
This work has been supported by the Fonds de la Recherche Scientifique (FRS-FNRS Belgium) through the PdR Grant No. T.0238.13 - AIXPHO, and by the European Research Council 
(Project SEED, grant agreement 320971). Work at Stony Brook was supported by US DOE grant No. DE-FG02-08ER46550.
Computational resources have been provided by the supercomputing facilities
of the Universit\'e catholique de Louvain (CISM/UCL)
and the Consortium des Equipements de Calcul Intensif en
F\'ed\'eration Wallonie Bruxelles (CECI) funded by the FRS-FNRS under Grant No. 2.5020.11.

\section*{APPENDIX. FIRST-PRINCIPLES CALCULATIONS : TECHNICAL DETAILS}

All first-principles calculations are done using ABINITv8.4.1 (main executable abinit as well as postprocessor ElectronPhononCoupling). The computations use
%within 
density functional theory (DFT, ground state and electronic structure) and 
density functional perturbation theory (DFPT, phonons and electron-phonon coupling),
as described in Refs.~\onlinecite{Gonze2009,Gonze2016}, 
with the PBE generalized gradient approximation (PBE-GGA) \cite{Perdew1996,Perdew1997} for MgO, 
and local density approximation (LDA) for LiF.
See Ref.~\onlinecite{Martin2004} for more information about the basic theories.
The effective masses have been computed directly using the scheme from Ref.~\onlinecite{Laflamme2016}.
The Allen-Heine-Cardona formalism is used for the computation
of the Fan and Debye-Waller self energies \cite{Allen1976,Allen1981,Ponce2014a,Antonius2015}.
The summation over unoccupied bands is handled efficiently thanks to a reformulation 
in terms of a Sternheimer equation \cite{Gonze2011}.
However, in order to treat correctly the dynamical self energy, the contribution from a number of ``active" bands is computed 
explicitly by a sum over states,
as described in Ref.~\onlinecite{Ponce2015}, see Eq.(15).

 Atomic masses from natural isotopic ratio have been used throughout, namely (in atomic mass units), 
 $M_{Mg}=24.305$, $M_{O}=15.9994$, $M_{Li}=6.941$ and $M_{F}=18.9984032$.

Norm-Conserving Pseudopotentials for MgO were taken from the ABINIT web site, generated using ONCVPSP \cite{Hamann2013}, while for LiF Troullier-Martins pseudopotentials have been used.
For both MgO and LiF, we used a 8$^3$
Monkhorst-Pack grid sampling of the Brillouin Zone for the electronic wavefunctions, and a 50 Ha kinetic energy plane-wave cut-off.
The summation over unoccupied bands for the dynamical self energy is made over 6 conduction bands for MgO, 
and 3 conduction bands for LiF, corresponding to a range of more than 10 eV above the CBM. 
The difference between the dynamical denominator (including the phonon frequency) and the static denominator
(with only the electronic eigenenergy differences) is smaller than 1\% above these bands, 
hence the summation over states can be safely replaced by the Sternheimer equation beyond them.
Note that the replacement of an explicit sum over states, that includes a physical infinitesimal imaginary
$\eta$, by a static contribution from the Sternheimer equation~\cite{Ponce2015}, destroys the 
Kramers-Kronig relations Eqs.~\ref{eq:ReSigma} and~\ref{eq:DerivSigma}. Of course, this has no
influence on the real and imaginary parts directly computed in 
the low-energy region of the  conduction bands or in the valence bands. 

\begin{figure}
\centering
\includegraphics[width=0.45\textwidth]{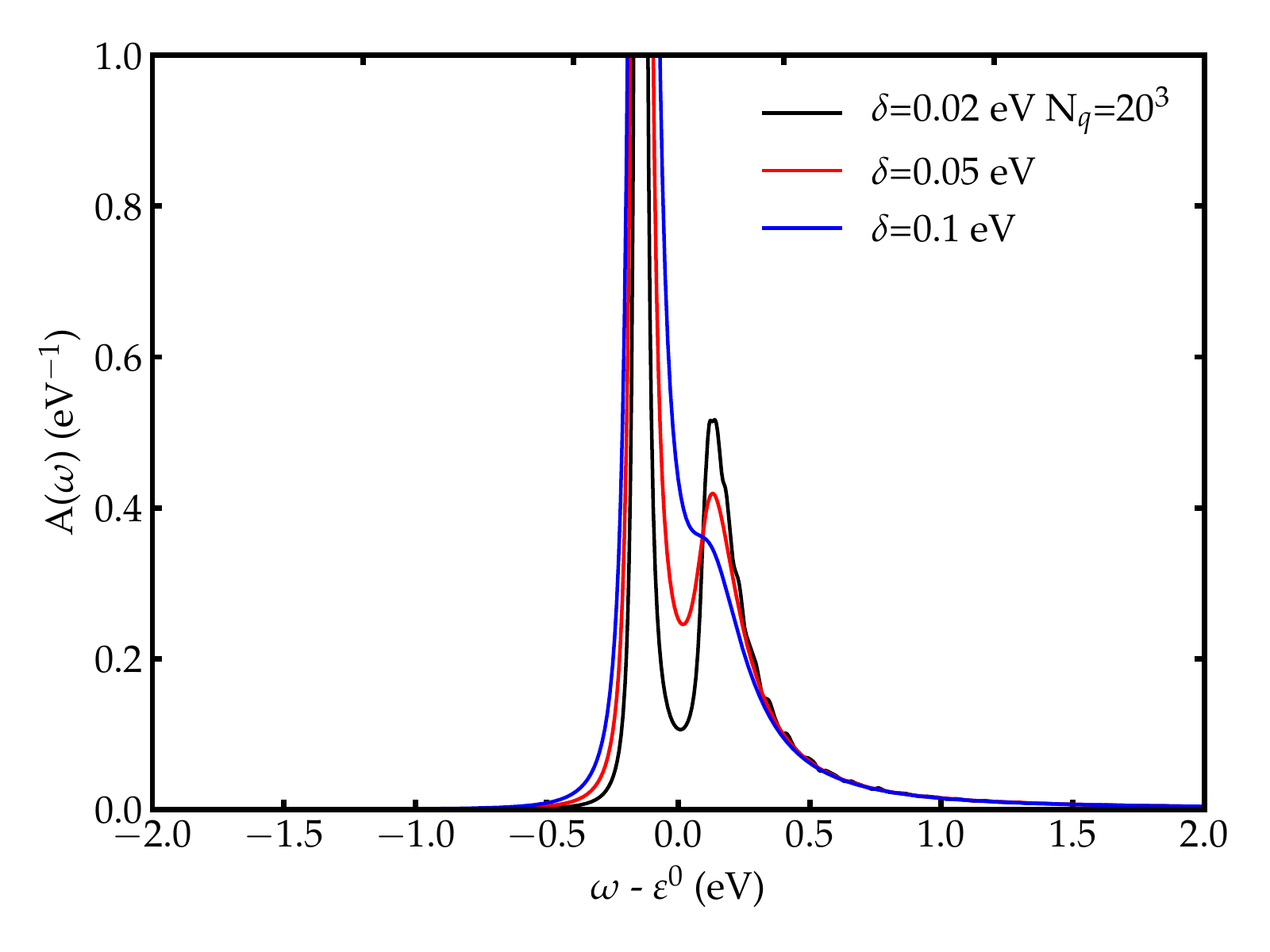}
\caption{Spectral function (Dyson-Migdal) of the MgO CBM self energy with a 20$^3$ q-grid, 
and decreasing $\delta$ = 0.1, 0.05, 0.02 eV. The separation between the quasiparticle peak 
and the satellite is still not complete with the smallest $\delta$ value. 
\label{fig:MgO_CBM_sfDM_delta_20x}
}
\end{figure}

\begin{table}[h]
\caption{\label{tab:selfenergy_convergence}MgO self energy (eV) and frequency derivative (adimensional) at $\omega=\varepsilon_{\mathbf{k}n}$,
for the conduction band minimum, for different wavevector samplings and broadening factors $\delta$ . 
}
\begin{tabular}{l | c | c | c}
 \hline \hline
Wavevector grid & $\delta$(eV) & $\Sigma$ & $\partial\Sigma/\partial\omega$ \\ 
20$^3$ & 0.01 & -0.1391 - i 0.0008 & -0.1603 - i 0.0089 \\
20$^3$ & 0.02 & -0.1390 - i 0.0032 & -0.1589 - i 0.0177 \\
20$^3$ & 0.05 & -0.1381 - i 0.0078 & -0.1498 - i 0.0421\\
20$^3$ & 0.10 & -0.1352 - i 0.0147 & -0.1232 - i 0.0712 \\
\hline
20$^3$ & 0.01 & -0.1391 - i 0.0016 & -0.1604 - i 0.0089 \\
32$^3$ & 0.01 & -0.1602 - i 0.0031 & -0.3086 - i 0.0294 \\
48$^3$ & 0.01 & -0.1746 - i 0.0050 & -0.4975 - i 0.0619 \\
96$^3$ & 0.01 & -0.1912 - i 0.0062 & -0.6264 - i 0.0968 \\
\hline \hline
\end{tabular}
\end{table}

The phonon wavevector sampling that is needed to get converged self energies and spectral functions 
is a delicate issue, already mentioned in Ref.~\onlinecite{Antonius2015}.
The most difficult case is the CBM of MgO, since this band disperses quite strongly, 
as can be deduced from its low effective mass. %Indeed, 
In order to get the imaginary part of the self energy, the wavevector sampling ought to sample %correctly the 
electronic eigenenergies such that their differences are not larger than
%with a characteristic value on the order of 
the LO phonon frequency. This is hard to achieve when 
%the curvature of the bands is large, hence, when 
the effective mass is small.  As described in Ref.~\onlinecite{Ponce2015}, see in particular Eq. (16),
a broadening factor $\delta$  is used to smooth the computed functions.
Too large a $\delta$ value prevents distinguishing characteristic features of the self energy 
or spectral function, which may typically appear at scales of the LO phonon frequency.
 
Fig. \ref{fig:MgO_CBM_sfDM_delta_20x} presents the D-M spectral function of the CBM of MgO 
with a 20$^3$ sampling, and %decreasing 
$\delta$ values 0.1, 0.05 eV and 0.02 eV.
Larger $\delta$ smoothes the function, but also %significantly 
modifies its shape: the
clear and physical separation between the quasiparticle peak and the satellite is %completely 
washed out.
 
The values of the self energy at the CBM are also strongly affected, 
as can be seen in Table \ref{tab:selfenergy_convergence}.  
From the numbers in this table, one can appreciate the convergence of the real part of the
self energy as the inverse of $N_q$, 
mentioned in Sec. VII, mathematically derived in Ref.~\onlinecite{Ponce2015}. This convergence study also highlights why the ZPR of the MgO CBM obtained in the
present work, namely -207meV, is quite different from the same quantity presented in 
Table I of Ref.~\onlinecite{Antonius2015}, $\Sigma^{dyn}(\epsilon^0)$=-143meV, although the same software and pseudopotentials have been used.

\begin{figure}
\centering
\includegraphics[width=0.45\textwidth]{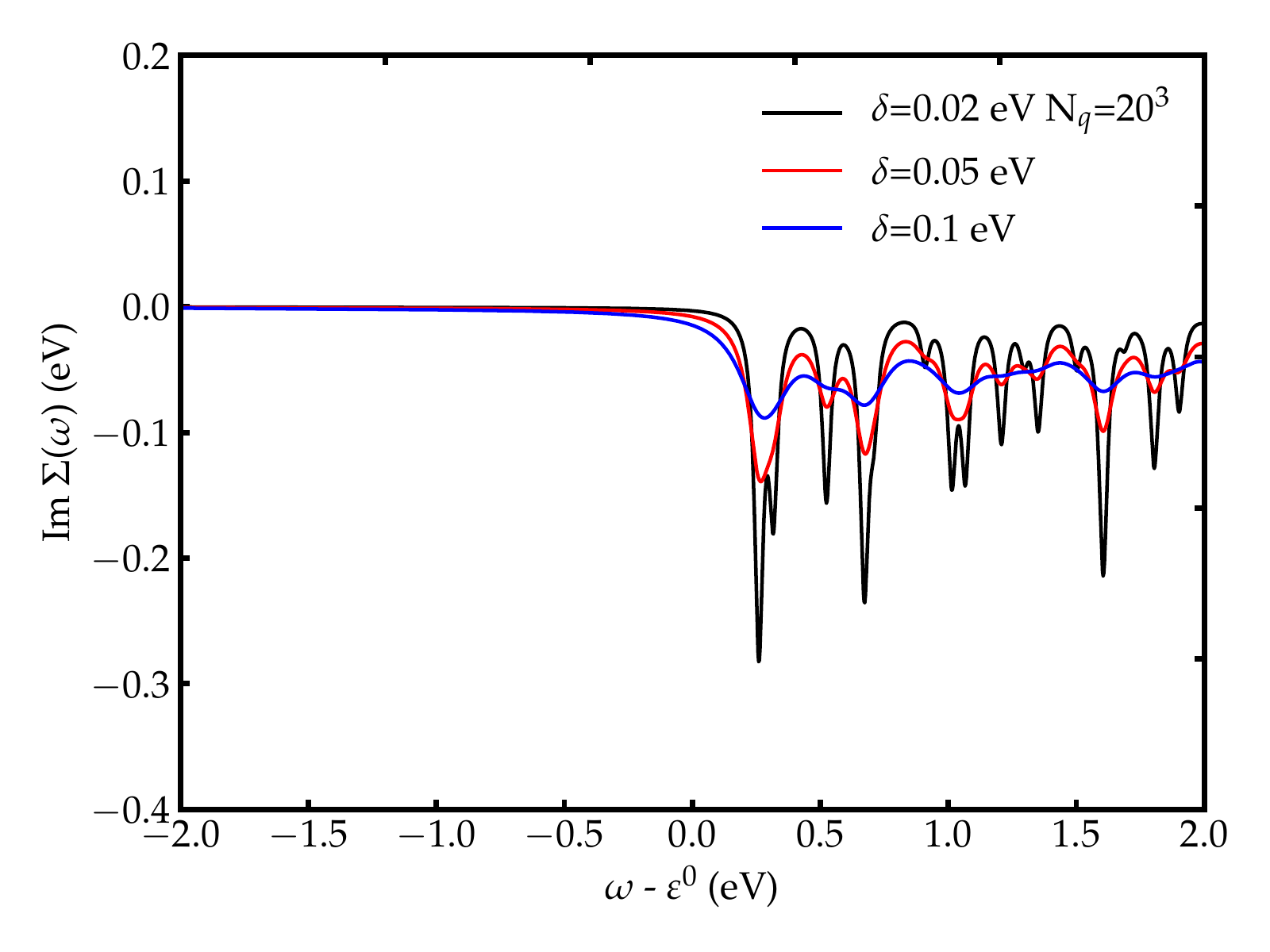}
\caption{The imaginary part of the MgO CBM self energy with a 20$^3$ q-grid 
and decreasing $\delta$ = 0.1, 0.05, 0.02 eV. 
\label{fig:MgO_CBM_slf_Im_delta_20x}
}
\end{figure}

\begin{figure}
\centering
\includegraphics[width=0.45\textwidth]{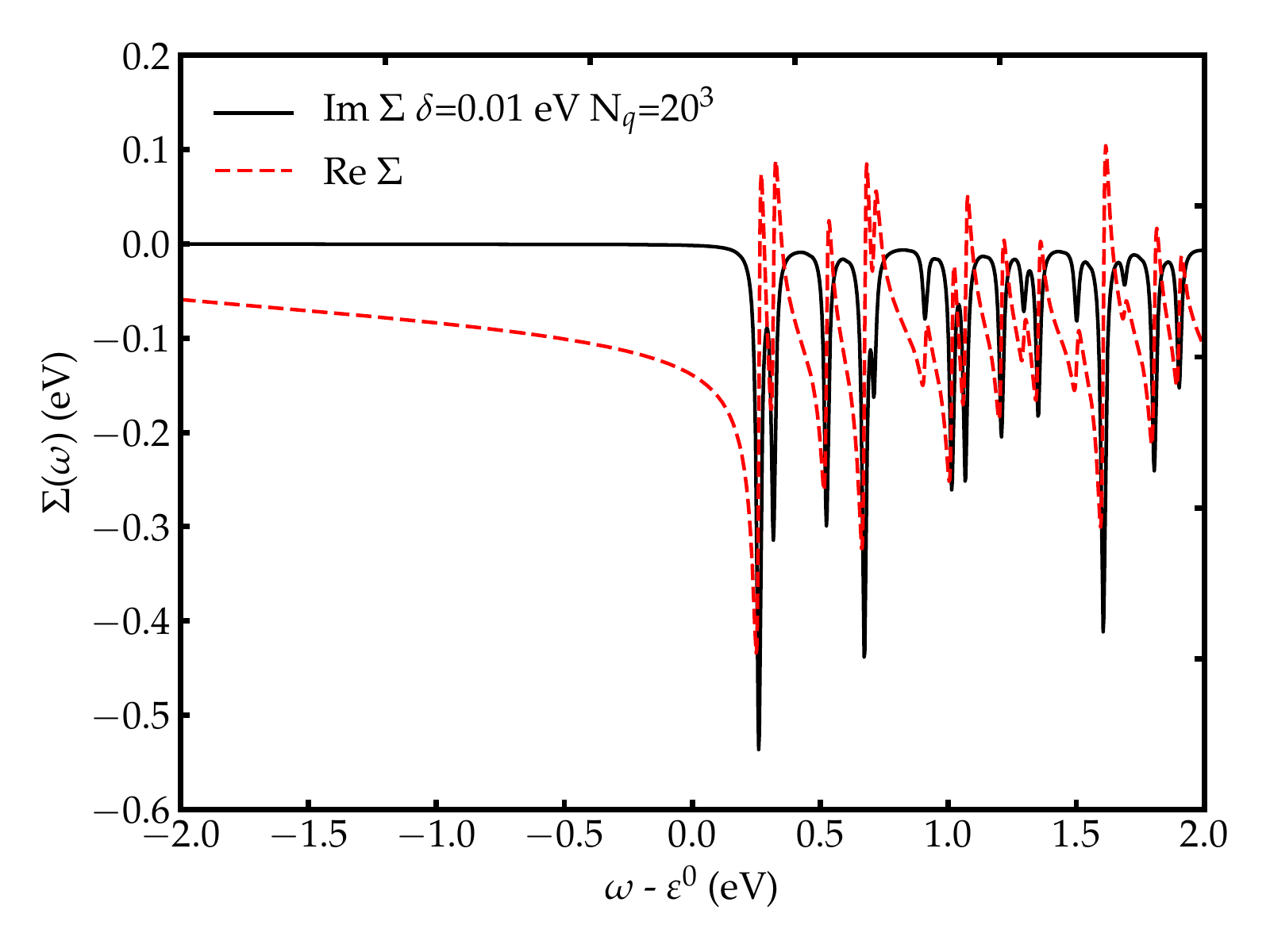}
\caption{The MgO CBM self energy with $\delta$ = 0.01 eV and a 20$^3$ q-grid. 
Full black line: imaginary part; 
dotted red line: real part. 
\label{fig:MgO_CBM_slf_001_20x}
}
\end{figure}

A 0.01 eV broadening appears  maximum %limit 
for a %reasonably 
meaningful investigation of the different spectral functions and self energies. 
%Actually, 
The self energy is even noisier than the spectral function 
(see  Fig. \ref{fig:MgO_CBM_slf_Im_delta_20x}).  %Of course, 
Going from a 20$^3$ sampling, that delivers the meaningless
Fig. \ref{fig:MgO_CBM_slf_001_20x}, to our best 96$^3$ grid, Fig. \ref{fig:MgO_CBM_slf_001_96x}, 
%with still some noise, means 
requires increasing by two orders of magnitude the computational effort,
%and also possibly meeting with 
introducing possible computer memory problems, and still leaves some noise.
%depending on the computer that is used.

%

\begin{figure}
\centering
\includegraphics[width=0.45\textwidth]{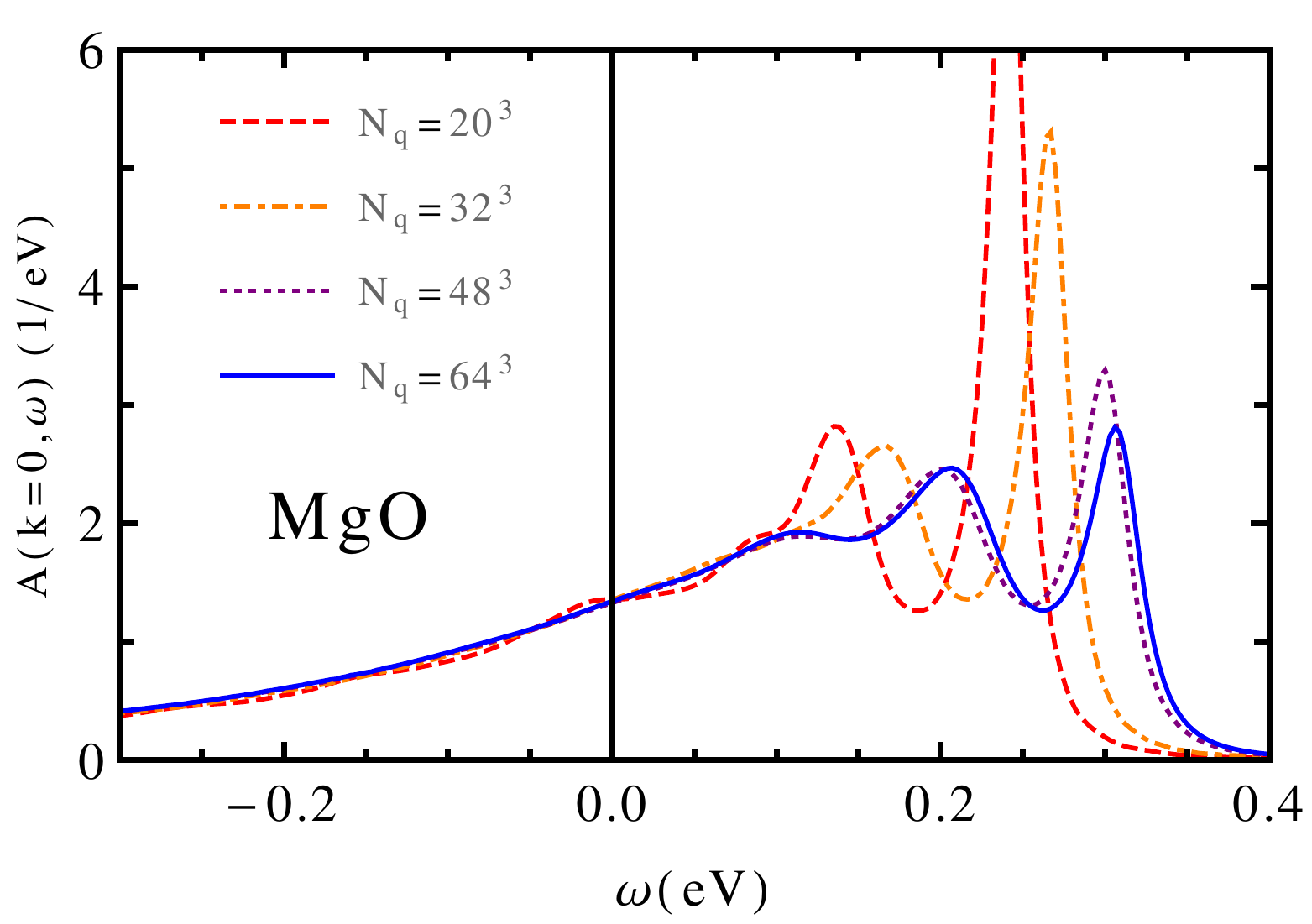}
\caption{Cumulant spectral function of the MgO VBM %self energy 
with $\delta$ = 0.01 eV and 20$^3$, 32$^3$, 48$^3$ and 64$^3$ q-grids. 
\label{fig:MgO_VBM_spCUM_001_20x32x48x64x_v3.pdf}
}
\end{figure}

The convergence parameters also impact the cumulant spectral functions, although they are %globally 
more easily converged than the
self energy or the D-M spectral function. The cumulant spectral functions corresponding 
to the VBM of MgO with $\delta$ = 0.01 eV and 20$^3$, 32$^3$, 48$^3$, and 64$^3$ grids 
are presented in Fig. \ref{fig:MgO_VBM_spCUM_001_20x32x48x64x_v3.pdf}.

%------------------------------------------------------------------------------

%\bibliographystyle{apsrev}
\bibliography{cumulant_v7.1}

\begin{thebibliography}{98}
\bibitem [{not({\natexlab{d}})}]{note-smalldelta}%
  \BibitemOpen
  \href@noop {} {}\bibinfo {note} {{Note however that the lack of structure at
  the broad maximum of the spectral function for large $\alpha$ is also a side
  effect of our numerical broadening of the self-energy, albeit quite small -
  of order 0.12$\omega_{LO}$. Indeed, at zero broadening, the spectral function
  is made of a sum of functions, each of which has one inverse-square
  divergence appearing at a multi-phonon threshold, as described at the end of
  Sec. IV. These divergencies are spaced by $\omega_{LO}$, corresponding to
  functions differing by one phonon. For such a function corresponding to $n$
  $LO$ phonons, the numerical broadening is $n$ times 0.12$\omega_{LO}$, and
  thus exceeds the spacing of the multiphonon peaks beyond 8 phonons, which
  happens at the maximum of the spectral functions beyond
  $\alpha=8$.}}\BibitemShut {Stop}%
\end{thebibliography}

\end{document}